\documentclass[aip, prl,twocolumn,superscriptaddress,nobibnotes,nofootinbib]{revtex4}
\setcitestyle{super}
\usepackage[utf8]{inputenc}
\usepackage{amsmath}
\usepackage[english]{babel}
\usepackage[dvipsnames]{xcolor}
\usepackage{amssymb}
\usepackage{textcomp}
\usepackage{graphicx}
\usepackage{textgreek}
\usepackage{latexsym}
\usepackage{braket}
\usepackage[nolist]{acronym}
\usepackage{siunitx}
\usepackage{braket}

\usepackage{upgreek}
\usepackage{xcolor}
\AtBeginDocument{\usepackage{booktabs}}
\usepackage[percent]{overpic}
\usepackage{adjustbox}

\usepackage{hyperref}
\hypersetup{
  colorlinks=true,
  linkcolor=black,
  citecolor = blue,
  urlcolor= black
}

\begin{document}
%


\title{Measurement of Gravitational Coupling between Millimeter-Sized Masses
}

\newcommand{\1}{Institute for Quantum Optics and Quantum Information (IQOQI) Vienna, Austrian Academy of Sciences, Boltzmanngasse 3, 1090 Vienna, Austria}
\newcommand{\2}{Vienna Center for Quantum Science and Technology (VCQ), Faculty of Physics, University of Vienna, Boltzmanngasse 5, 1090 Vienna, Austria}
\newcommand{\3}{Research Platform TURIS, University of Vienna, Boltzmanngasse 5, 1090 Vienna, Austria}
\newcommand{\+}{These authors contributed equally to this work}

\author{Tobias Westphal}
\affiliation{\1}
\author{Hans Hepach}
\thanks{These authors contributed equally to this work}
\affiliation{\1}
\author{Jeremias Pfaff}
\thanks{These authors contributed equally to this work}
\affiliation{\2}
\author{Markus Aspelmeyer}
\affiliation{\1}
\affiliation{\2}
\affiliation{\3}

\begin{abstract} 
Gravity is the weakest of all known fundamental forces and continues to pose some of the most
outstanding open problems to modern physics: it remains resistant to unification within the standard
model of physics and its underlying concepts appear to be fundamentally disconnected from quantum
theory\cite{Unruh1984,Preskill1992,Greenberger2010a,Penrose2014}. Testing gravity on all scales is therefore an important experimental endeavour\cite{Will2006,Adelberger2001,Hossenfelder}. 
Thus far, these tests involve mainly macroscopic masses on the kg-scale and beyond\cite{Gillies2014}. Here we show gravitational coupling between two gold spheres of 1 mm radius, thereby entering the regime of sub-100 mg sources of gravity. Periodic modulation of the source mass position allows us to perform a spatial mapping of the gravitational force. Both linear and quadratic coupling are observed as a consequence of the nonlinearity of the gravitational potential. Our results extend the parameter space of gravity measurements to small single source masses and small gravitational field strengths.
Further improvements will enable the isolation of gravity as a coupling force for objects below the Planck mass. This opens the way to a yet unexplored frontier of microscopic source masses, which enables new searches of fundamental interactions\cite{Feldman2006,Burrage2015,Hamilton2015} and provides a natural path towards exploring the quantum nature of gravity\cite{DeWitt2011,Bose2017,Marletto2017,Belenchia2018}.
\end{abstract}
\maketitle
%

The last decades have seen numerous experimental confirmations of Einstein's theory of relativity, our best working theory of gravity, by observing massive astronomical objects and their dynamics\cite{Will2006,Ciufolini2004,Ransom2014}. This culminated in the recent direct detection of gravitational waves from the merger of two black holes\cite{Abbott2016}, and the direct imaging of a supermassive black hole\cite{Akiyama2019}. During the same time, Earth-bound experiments have continuously been increasing their sensitivity for gravity phenomena on laboratory scales, including for general relativistic effects\cite{Chou2010,Asenbaum2017}, tests of the equivalence principle\cite{Ritter1990,Adelberger2001,Rosi2017}, precision measurements of Newton's constant\cite{Gundlach2000,Quinn2013,Rosi2014} or tests of the validity of Newton's law at $\mathrm{\upmu m}$-scale distances\cite{Geraci2008,Tan2020,Lee2020}. While test masses in such experiments span the whole range from macroscopic objects to individual quantum systems\cite{Colella1975,Nesvizhevsky2002,Chou2010,Asenbaum2017} the gravitational source is typically either Earth or masses on the kg-scale and beyond\cite{Gillies2014}. A yet unexplored frontier is the regime of microscopic source masses, which enables new searches of fundamental interactions\cite{Feldman2006,Burrage2015,Hamilton2015} and provides a natural path towards exploring the quantum nature of gravity\cite{DeWitt2011,Bose2017,Marletto2017,Belenchia2018}. Here we show gravitational coupling between two gold spheres of 1\,mm radius, thereby entering the regime of sub-100\,mg sources of gravity.

Experiments with smaller source masses are excessively more difficult -- the gravitational force generated at a given distance by a spherical mass of radius $R$ shrinks with $R^{-3}$ -- and hence only few experiments to date have observed gravitational signatures of gram-scale mass configurations\cite{Hoskins1985,Mitrofanov1988,Lee2020}. In one case, a hole-pattern in a rotating, 5\,cm diameter attractor disk made from platinum generated a periodic mass modulation of a few hundred mg, which was resolved in a torsional balance measurement\cite{Lee2020}. In another case, a single 700 mg tungsten sphere was used to resonantly excite a torsional pendulum\cite{Mitrofanov1988}. Isolating gravitational interactions generated by even smaller, single source-mass objects is a challenging task as it requires increasing efforts to shield residual contributions from other sources of acceleration, in particular of seismic and electromagnetic nature\cite{Schmole2016}. Also, resonant detection schemes, which are typically employed to amplify the signal above readout noise, amplify displacement noise as well and hence don't yield any gain in terms of separating the signal from other force noise sources. We are overcoming this limitation by combining time-dependent gravitational accelerations with an off-resonant detection scheme of a well-balanced differential mechanical mode and independent noise estimation. 

\section*{Experiment}

\begin{figure}[htb]
    \includegraphics[width=0.9\linewidth]{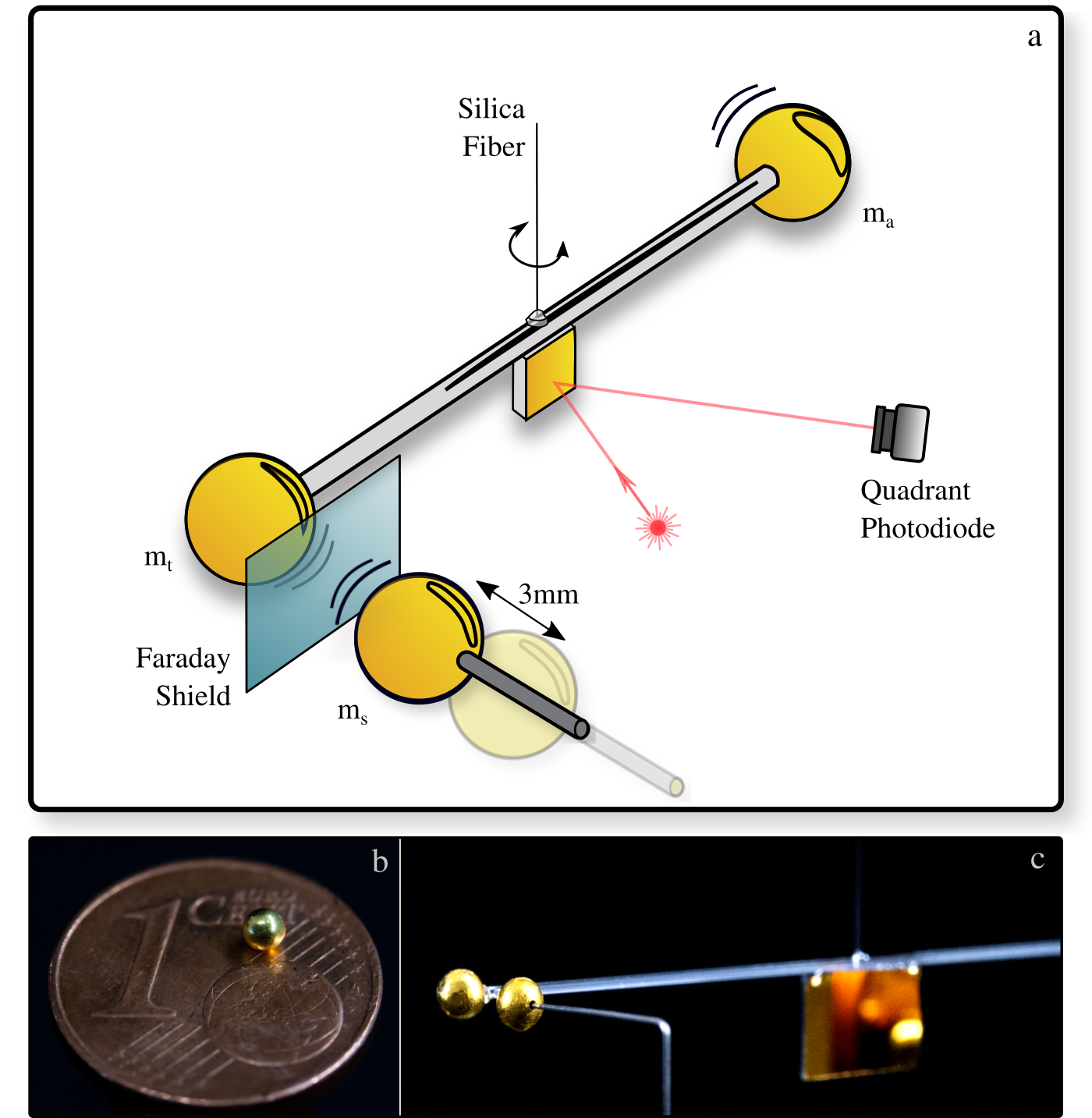}
    \caption[Experimental setup]{
    (a) Schematic of the experiment. 
    The torsion pendulum, which acts as transducer for gravitational acceleration, consists of two $r\approx1\,\mathrm{mm}$ gold spheres held at $40\,\mathrm{mm}$ center distance by a glass capillary. 
    One mass serves as $m_t=90.7\,\mathrm{mg}$ test mass, the other as $m_a=91.5\,\mathrm{mg}$ counterbalance that provides vibrational noise common mode rejection. 
    A $4\,\mathrm{\mu m}$ diameter silica fiber provides a soft $f_0\approx3.6\,\mathrm{mHz}$ torsional resonance separated well from other degrees of freedom. 
    The torsion angle is read out via an optical lever directed to a quadrant photodiode. 
    The gravitational interaction is modulated by harmonically moving the $m_s=92.1\,\mathrm{mg}$ source mass by $\approx 3$mm at a frequency $f_{mod}=12.7\,\mathrm{mHz}$ well above the torsion resonance using a shear piezo. 
    Direct electrostatic coupling is suppressed by discharging and a $150\,\mathrm{\mu m}$ thick Faraday shield.
    (b) Source mass on a 1 Euro Cent coin.
    (c) Photo of the torsion pendulum and the mounted source mass.}
    \label{fig:schematic}
\end{figure}

\begin{figure}[!t]
\centering
\includegraphics[width=1.0\linewidth]{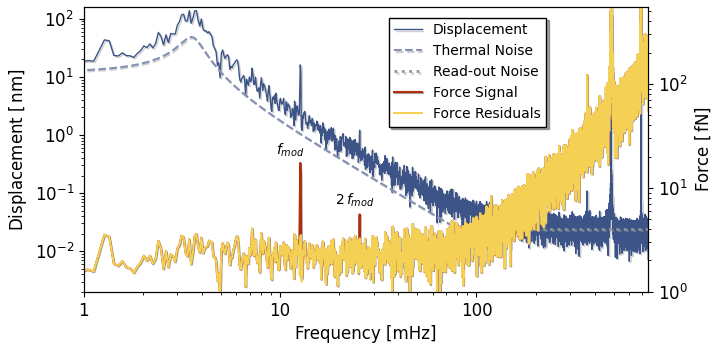}
    \caption[Spectrum of a measurement run]{The rotation of the torsional oscillator measured over a period of 13.5 hours is calibrated as test mass displacement (blue) and respective exerted force (light red). 
    The measurement is limited by both thermal noise (dashed) and} diurnally varying white force noise below $100\,\mathrm{mHz}$, while white displacement readout noise (dotted) dominates above. The torsional motion is decoupled well from other oscillator modes, starting at 0.5\,Hz.
    Force residuals (yellow) represent the difference between measured and expected gravitational force. The spectrum shows gravitational coupling at $f_{mod}=12.7\,$mHz modulation frequency plus the nonlinearity of the gravitational potential at $2f_{mod}=25.4\,$mHz well above the oscillator's resonance frequency $f_0=3.59\,\textrm{mHz}$. The flatness of the force residuals indicate that both signals are in agreement with the expected gravitational signal. Other sources of nonlinearity are found to be negligible\cite{SI}. 
    \label{fig:spectrum}
\end{figure}

In our experiment, the gravitational source is a nearly spherical gold mass of radius $r=(1.07\pm 0.04)\,\mathrm{mm}$ and mass $m_s=(92.1\pm 0.1)\,\mathrm{mg}$. A similarly sized gold sphere acts as test mass with $m_t=(90.7\pm 0.1)\,\mathrm{mg}$. The idea is that a periodic modulation of the source mass position generates a time-dependent gravitational potential at the location of the test mass, whose acceleration is measured in a miniature torsion pendulum configuration (Figure~\ref{fig:schematic}).  The experiment is conducted in high vacuum ($6\times 10^{-7}\,\mathrm{mbar}$), which minimizes residual noise from acoustic coupling and momentum transfer of gas molecules\cite{Schmole2016,SI}. To prevent non-linear coupling of high-frequency vibrations into the relevant low-frequency measurement band around the modulation frequency $f_{mod}=12.7\,\mathrm{mHz}$, the pendulum support structure is resting on soft, vacuum-compatible rubber feet\cite{Shimoda2019}.\\ 
We optically monitor the angular deflection of the pendulum, which provides a calibrated readout of the test mass motion with a detector-noise limited sensitivity of $\approx 2\,\mathrm{nm}/\sqrt{\mathrm{Hz}}$. Figure \ref{fig:spectrum} shows an amplitude spectrum of the test mass displacement. The torsional mode resembles a damped harmonic oscillator with resonance frequency $f_0=3.59\,\textrm{mHz}$ and mechanical quality factor $Q=4.9$. It is well decoupled from other oscillation modes of the pendulum, which do not occur below $0.5\,\mathrm{Hz}$. Diurnal variations in the low-frequency noise-floor limit the times of best sensitivity to those hours during the night where local public transport and pedestrian- and car-traffic are minimized (typically between midnight and 5 a.m.). There, the test mass oscillation was mainly governed by thermal noise\cite{SI}. The corresponding force spectrum was obtained by deconvolution of the test mass displacement time series with the inverse of the deduced mechanical susceptibility. It exhibits a flat noise floor that allows off-resonant detection of test mass accelerations at frequencies up to $0.1\,\mathrm{Hz}$ at a sensitivity better than $2\times 10^{-11}\,\mathrm{m}/\mathrm{s}^2$ within half a day. We use this to obtain a live-estimate of the pendulum noise conditions, which is combined with inverse variance weighting of data to optimize the information obtained per experimental run.\\

The source mass is mounted on a $300\,\mathrm{\mu m}$ diameter titanium rod that is connected to a bending piezo, which provides more than $5\,\mathrm{mm}$ travel range in the vicinity of the test mass. The geometric separation of the masses was determined with an accuracy of $20\,\mathrm{\mu m}$ by linking the drive-piezo signal to position information from video tracking through a template matching algorithm\cite{SI}. During the experiment, the center distance between source and test mass was varied between $2.5\,\mathrm{mm}$ and $5.8\,\mathrm{mm}$, with a minimal surface distance of $\approx0.4\,\mathrm{mm}$.\\ 
To isolate gravity as a coupling force we need to minimize other influences on the test mass. In addition to seismic and acoustic effects these are predominantly electromagnetic interactions. We ground the source mass by directly connecting it to the vacuum chamber. The test mass is discharged to less than $8\times10^4$ elementary charges using ionized nitrogen\cite{Ugolini2011}, a method that was developed for charge mitigation in interferometric gravitational wave detectors\cite{Abbott2016}. Further shielding is provided by a $150\,\mathrm{\mu m}$ thick conductive Faraday shield of gold-plated aluminum that is mounted between source- and test mass. In that way, electrostatic forces were suppressed to well below $3\,\%$ of the expected gravitational coupling strength. Other shielding measures include housing the source-mass drive piezo inside a Faraday shield to suppress coupling via the applied electric fields, as well as gold-coating and grounding most surfaces inside the vacuum chamber to exclude excitation from time-varying charge distributions. We also rule out the presence of relevant magnetic forces by independently measuring the magnetic permeabilities of the masses. As expected, permanent magnetic dipoles are negligible in both (diamagnetic) gold and (paramagnetic) titanium. Coupling via induced magnetic moments, either from Earth's magnetic field or from low-frequency magnetic noise originating in nearby urban tram traffic, is also found to be orders of magnitude below the expected gravitational coupling strength. In the current experimental geometry, Casimir forces are negligible, although they will likely become a dominant factor for significantly smaller masses\cite{Canaguier-Durand2011}. 
Another relevant noise source is Newtonian noise, which is caused by non-stationary environmental gravitational sources and cannot be shielded. For comparison, at a center separation of $2.5\,\mathrm{mm}$ the static gravitational force between our masses is expected to be $9\times 10^{-14}\,\mathrm{N}$. The same gravitational force is exerted on the test mass by a human experimenter standing at a distance of $2.5\,\mathrm{m}$, or by a typical Vienna tram at $50\,\mathrm{m}$ distance from the laboratory building. Consequently, our experiment experiences a complex low-frequency gravitational noise of urban origin. It is obvious that such gravitational noise sources will pose an increasing challenge for future experiments. At present, our torsion pendulum is sufficiently small compared to the distance of typical environmental sources to be insensitive due to common mode rejection. 
\section*{Results}

\begin{figure}[!ht]
    \centering
    \includegraphics[width=1.0\linewidth]{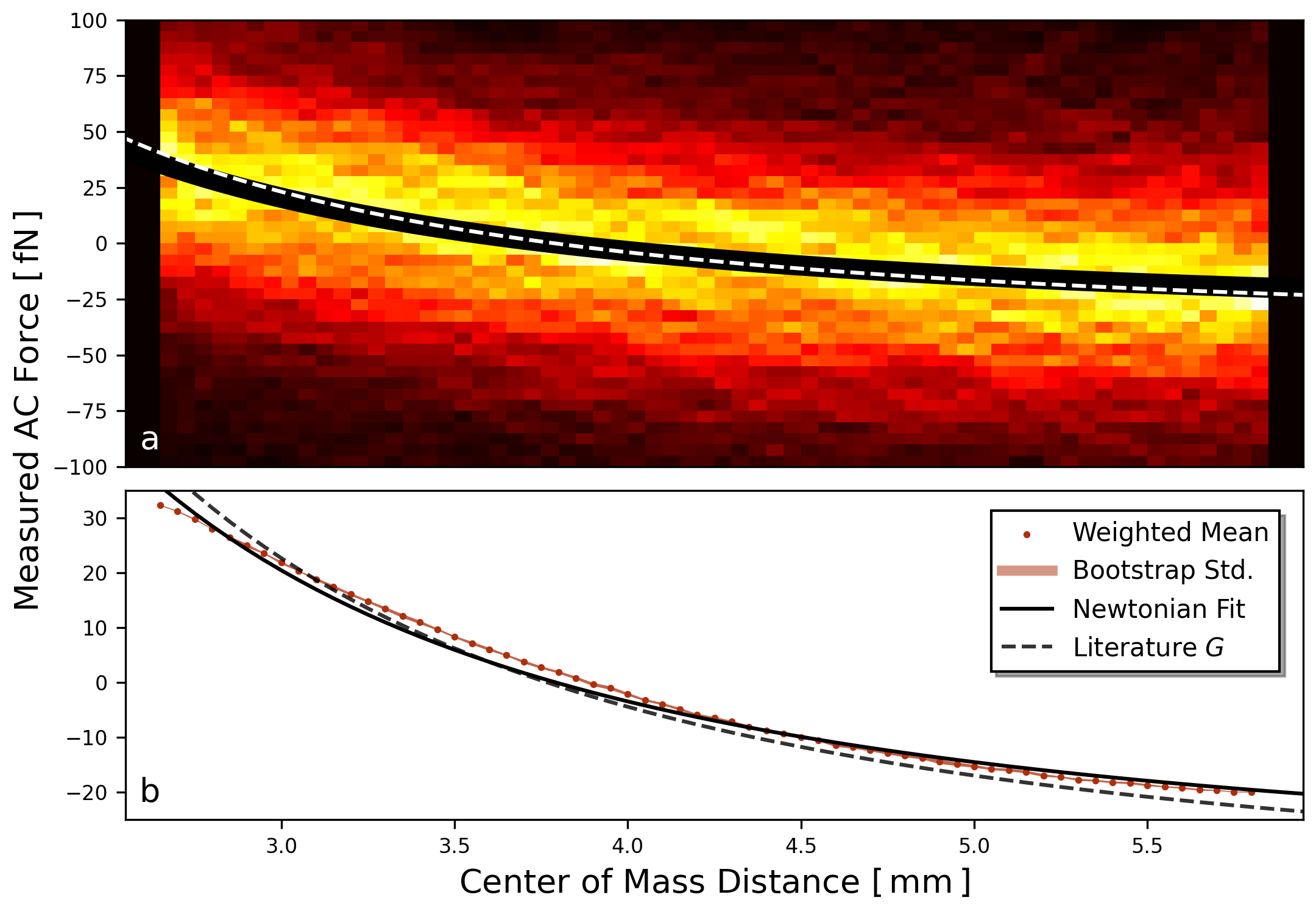}
    \caption[Probability density of F(d)]{The probability density distribution of exerted force versus source- test mass separation (a) shows the spatial non-linearity of the source-mass potential. This method makes full use of our measurement data by taking into account weighting with the current noise conditions. The apparent width of the distribution is mainly given by noise at frequencies other than the modulation. (b) The actual measurement precision becomes apparent by extracting the mean of the force measured at given distances (red points) along with their standard deviation obtained by bootstrapping (red band). Note that the small statistical error collapses the error band to the narrow red line. A full weighted fit of the 13.5\,h long data segment with Newtonian gravity (black line) yields $G_{fit}=(5.89\pm0.20)\times10^{-11}\,\mathrm{N}\,\mathrm{m}^2/\mathrm{kg}^2$ which is in agreement with our long-term combined value (see Figure~\ref{fig:G}). Comparison with the literature (CODATA) value $G_{CODATA}=6.67\times10^{-11}\,\mathrm{N}\,\mathrm{m}^2/\mathrm{kg}^2$ (dashed line) and our known systematic errors shows that influences other than gravity are below $10\,\%$.}
    \label{fig:bonfire}
\end{figure}

We observe gravitational coupling between the two masses by harmonically modulating the source mass position $x_s=d_0+d_m\,\cos(2\pi f_{mod}t)$ at a frequency $f_{mod}=12.7\,\mathrm{mHz}$, well above the fundamental torsional pendulum resonance (mean center-of-mass distance $d_0$; modulation amplitude $d_m$). In this frequency regime the test mass response becomes independent of the pendulum properties and behaves essentially as a free mass. According to Newton's law the source mass generates a gravitational acceleration of the test mass of $a_G=G\frac{m_s}{x_s(t)^2}$
($G$: Newton's constant). The 1/r dependence of the gravitational potential results in higher-order contributions to the coupling at multiples of the modulation frequency $f_{mod}$. Figure \ref{fig:spectrum} shows the measured force spectrum of the test mass for a distance modulation of $d_m\approx1.6\,\mathrm{mm}$ at a center separation of $d_0\approx 4.2$mm, in which we resolve the linear and quadratic acceleration modulations at $f_{mod}$ and $2f_{mod}$, respectively. The flat noise residuals clearly indicate that the observed peak heights 
agree well with the expected force modulation due to Newtonian gravity. This confirms the gravitational origin of the interaction. \\
To more accurately quantify the strength of the coupling we correlate the measured separation $x_s$ between source and test mass with the independently inferred force on the test mass. This provides us with a position-dependent mapping of the gravitational force (Figure \ref{fig:bonfire}). All data evaluation is carried out in post-processing and using non-causal zero-phase filtering\cite{SI}. Each data set, which consists of up to 13.7 hours long measurements, is fitted using the expression for Newtonian gravity 
to obtain a value for the measured coupling strength.

\begin{figure}[!t]
\centering
\includegraphics[width=1.0\linewidth]{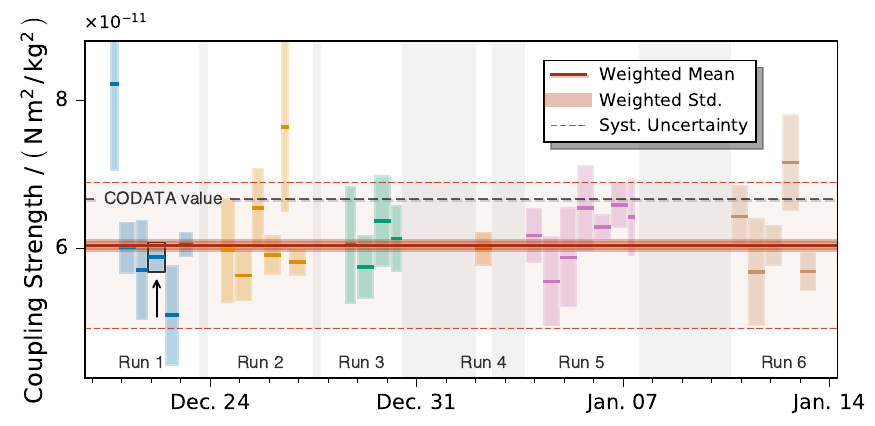}
\caption[Collection of measurement runs]{Over a stretch of three weeks during the seismically quiet Christmas season several measurement runs were conducted, with electrostatic coupling being excluded during gray periods. The dataset presented in Figure~\ref{fig:spectrum} and \ref{fig:bonfire} is highlighted in black (run 1). The statistical error ($1\sigma$ error bar) of the best fitting Newton-like coupling of each data segment (dark lines) varies due to highly non-stationary noise in the measurement band. 
Combining individual measurements weighted by their respective data quality yields $G_{comb} = (6.04\pm0.06)\times10^{-11}\,\mathrm{N}\,\mathrm{m}^2/\mathrm{kg}^2$. The plotted systematic uncertainty (red dashed) includes the identified systematic influences from the experiment as summarized in Table II. We explicitly do not claim a significant systematic deviation of the combined coupling value from the CODATA recommended value (blue dashed). The experiment and its data evaluation was mainly designed with the reduction of statistical errors in mind, and the deviation is fully covered by the known experimental systematics.}
    \label{fig:G}
\end{figure}

Figure \ref{fig:G} summarizes the results of a series of measurements that were taken during the seismically quiet Christmas season. The weighted combination of measurement runs yields a coupling constant of $G_{comb} = (6.04\pm 0.06) \times 10^{-11}\,\mathrm{m}^3\mathrm{kg}^{-1}\mathrm{s}^{-2}$, with a statistical uncertainty of $1\,\%$. The observed coupling deviates from the recommended CODATA value for Newton's constant ($G_{CODATA}=6.67430(15)\times 10^{-11}\,\mathrm{m}^3\mathrm{kg}^{-1}\mathrm{s}^{-2}$) by around $9\,\%$. This offset is fully covered by the known systematic uncertainties in our experiment, which include unwanted electrostatic, magnetic and gravitational influences from the masses and supports as well as geometric uncertainties in the center of mass distance due to the actual shape of the masses (we provide a detailed list in the Methods section\cite{SI}). Our results show that we are able to isolate gravity of a single, small source mass, with other influences being below the $10\,\%$ level. The small statistical error underlines the precision character of our measurement and our ability to measure even smaller source masses with our approach in the future.

\section*{Discussion and Outlook}
We have demonstrated gravitational coupling between a test mass and a $90\,\mathrm{mg}$ spherical source mass. To our knowledge this is the smallest single object whose gravitational field has been measured. Our measurements resolve an acceleration modulation of $\approx3\times10^{-10}\,\textrm{m}/\textrm{s}^2$ at the drive frequency $f_{mod}$ with high accuracy, quantified by a systematic uncertainty of $\approx3\times10^{-11}\,\textrm{m}/\textrm{s}^2$, and high precision, quantified by a statistical uncertainty of $\approx3\times10^{-12}\,\textrm{m}/\textrm{s}^2$ over an integration time of 350\,h. 
Contributions of non-gravitational forces could be kept below 10\,\% of the observed signal. 
Our results extend gravity experiments to the regime of small gravitational source masses, with the potential of going significantly smaller. As our current sensitivity is limited by the thermal noise floor of the torsion pendulum when operated under optimal ambient conditions, it can be readily improved by increasing the mechanical quality factor $Q$. Probing gravity of objects even smaller than the Planck mass $m_P\approx 22\,\mathrm{\mu g}=2.4\times 10^{-4}m_s$ should become possible for $Q\gtrsim20,000$\cite{SI}.
The thermal noise limited force sensitivity of these miniature torsion pendula can be improved even further by dissipation dilution as recently demonstrated\cite{Komori2020}. Another interesting route is to use levitated test masses, which provide almost perfect charge mitigation and similar acceleration sensitivities\cite{Prat-Camps2017,Timberlake2019,Monteiro2020,Lewandowski2020,Kawasaki2020}. To make full use of improvements in thermal noise it is important to understand and to mitigate other, predominantly anthropogenic, low-frequency noise sources that currently dominate our measurement performance most of the time. Possible countermeasures include proper choice of a remote laboratory location or pushing the experiment to much higher pendulum frequencies\cite{Schmole2016,liu2020prospects}. 
Going to smaller masses will involve additional challenges as other noise contributions will play an increasing role. For example, even for the ideal case of electrically neutral source and test mass, Casimir forces will become relevant at distances below $100\,\mathrm{\mu m}$. Note however, that electromagnetic shielding combined with a signal modulation can overcome this particular challenge. We also note that our method allowed us to directly measure the position-dependence of the gravitational force, which constitutes a 1D-mapping of the gravitational field strength. A more complex test mass geometry may enable a full mapping of the metric tensor of such small source masses\cite{Obukhov2019}.\\ 
Our experiment provides a viable path to enter and explore a new regime of gravitational physics that involves precision tests of gravity with isolated microscopic source masses at or below the Planck mass. This opens up new possibilities. For example, such measurements offer a different approach to determine Newton’s constant\cite{Schmole2016}, which to date remains the least well determined of the fundamental constants\cite{Speake2014}. In general, miniaturized precision experiments may allow tests of the inverse square law of gravity at significantly smaller scales than possible today (the best reported value is on the $50\,\mathrm{\mu m}$-scale\cite{Lee2020}), thereby probing, for example, fundamental features of string theory such as extra dimensions and massless scalar particles\cite{Adelberger2003}. Small source masses also allow to test consequences of new speculative scalar fields that have been discussed in the context of dark energy\cite{Feldman2006,Burrage2015,Hamilton2015}. More stringent parameter constraints for so-called Chameleon forces may already be possible for source masses at the level investigated here\cite{Feldman2006}. Another example is experimental access to gravitational accelerations at or even below the level of galaxy rotations. It has been suggested to modify Newtonian dynamics in this regime as an alternative to dark matter scenarios\cite{Milgrom1983}, and some of these models can therefore be directly tested\cite{Ignatiev2015}. Finally, the ongoing discussion on the quantum nature of gravity has revived a gedankenexperiment by Feynman to probe gravitational coupling between quantum systems\cite{DeWitt2011,Bose2017,Marletto2017}. The underlying question is whether it is possible to probe gravitational phenomena that cannot be explained by a purely classical source mass configuration\cite{Unruh1984}. Such experimental tests have thus far been elusive, since they require the ability to prepare quantum states of motion of the source mass. Since decoherence phenomena scale dramatically with system size, isolating gravity of microscopic masses is a necessary prerequisite for such future experiments. Our result provides a first step in this direction -- although we stress that, with current quantum experiments using sub-micron sized objects of $10^{-17}\,\mathrm{kg}$\cite{Tebbenjohanns2020,Delic2020}, adding quantum coherence to experiments in the relevant mass regime is yet a completely different experimental challenge.
%


\textbf{Acknowledgements} We thank Eric Adelberger, Aaron Buikema, Peter Graham, Nikolai Kiesel, Norbert Klein, Davide Racco and Jonas Schm\"ole for stimulating discussions. We are grateful for the suspension fiber provided by Arno Rauschenbeutel and Thomas Hoinkes and for the exceptional mechanical design assistance by Mathias Dragosits. This project was supported by the European Research Council (ERC) under the European Union’s Horizon 2020 research and innovation program (Grant Agreement No. 649008, ERC CoG QLev4G), by the Austrian Academy of Sciences through the Innovationsfonds Forschung, Wissenschaft und Gesellschaft, by the Alexander von Humboldt foundation through a Feodory Lynen fellowship (T.W.) and by the Austrian Federal Ministry of Education, Science and Research (project VCQ HRSM).  



\bibliographystyle{bibliography_nature}
\bibliography{milliG2020References_20200921_0046h}
\section{\large{Supplementary Material}}
\section{Torsion pendulum}
\subsection*{Geometry}
The torsion pendulum consists of the test mass ($m_t=90.7\,\mathrm{mg}$) and a counterbalance mass ($m_a=91.5\,\mathrm{mg}$). The masses are rigidly connected by a $37.9\,\mathrm{mm}$ long, square-shaped glass capillary with an outer dimension of $0.4\,\mathrm{mm}$ and $0.1\,\mathrm{mm}$ wall thickness ($9.6\,\mathrm{mg}$). It is supported by a $35\,\mathrm{mm}$ long quartz fiber pulled down to $3.6\,\mathrm{\mu m}$ diameter over a $20\,\mathrm{mm}$ section at the center for maximal torsional compliance. The $5-10\,\mathrm{mm}$ long, up to $50\,\mu\mathrm{m}$ wide taper region below this torsion section increases pitch- and roll-stiffness for better mode separation and provides a bigger cross section for attachment to the capillary. The attachment point was tuned to minimize static roll to $<10\,\mathrm{mrad}$. A $5.85\,\mathrm{mm}$ wide, $0.3\,\mathrm{mm}$ thin, hand ground and gold coated mirror ($27.0\,\mathrm{mg}$) was attached underneath the capillary for deflection readout of the torsion (yaw) angle. This turned out to point downwards by $50\,\mathrm{mrad}$ in pitch, potentially coupling non-torsion motion, in particular roll, into the yaw readout as well. All connections were established by UV-curing adhesive. 
The fundamental mode frequency $f_0$ is determined by the stiffness of the suspension fiber and the effective mass $m_{eff}=183.8\,\mathrm{mg}\approx m_t+m_a=182.2\,\mathrm{mg}$. The effective mass is almost minimal, i.e.\ hugely dominated by the two gold masses for the balanced, differential mode used here. It can only be reduced further by using a single mass in a non-differential setup, which would then, however, be subject to the full environmental translational acceleration.


\subsection*{Pendulum characterization}
Since we aim to characterize the gravitational force exerted by our source mass, we specify the test mass motion by its spatial displacement rather than by the pendulum deflection angle. The amplitude spectrum of the test mass displacement (shown in Figure \ref{fig:spectrum}) follows the slope of the mechanical susceptibility of a damped harmonic oscillator\footnote{Due to the low frequency of observation and the extremely good frequency separation of mechanical modes of a torsion pendulum in general, all transfer functions except the mechanical susceptibility of the yaw mode can be neglected.} which is solely defined by its eigenfrequency $f_0$, mechanical quality factor $Q$, and mass $m_{eff}$. The first two were obtained by a fit to the spectrum to be $f_0=3.59\,\textrm{mHz}$ and $Q=4.9$. Both parameters were verified during measurement runs and independently by a ringdown measurement with relatively small amplitude. For larger amplitudes, we observe a significant decrease in mechanical losses. Values in excess of $Q=2\times10^4$ have been observed for larger amplitudes ($\approx\pi$\,rad) and different rest positions. The effective mass has been determined by weighing and modelling all components except for the little amounts of glue that have been used.
        
The drive frequency $f_{mod}$ was chosen to lie well above the torsion resonance ($\approx3.6\,\textrm{mHz}$), yet far away from other, highly excited modes (e.g.\ around 500 and 700\,mHz) and sufficiently below the readout noise ($\approx$100\,mHz) such that the $2f_{mod}$ (and eventually the $3f_{mod}$) signal could be resolved. The intention of the exact choice of 12.7\,mHz to be such an odd number is to prevent frequency mixing occurring in nonlinear systems such as discretization, clipping etc.\ of narrow band signals to accidentally match with the gravitational signal.
        
By choosing the signal frequency in such a way, the test mass effectively behaves like a free mass ($m_{eff}$)\footnote{More precisely, if the signal frequency is well above the mechanical linewidth $\gamma_0/2\pi=f_0/Q$, i.e. $f_{mod}\gg \gamma_0/2\pi\approx0.7\,\textrm{mHz}$}. Hence, in this regime the response is mostly independent of the resonant enhancement factor $Q$ and the resonance frequency $f_0$.

\subsection*{Prospects towards smaller masses}
The experiment presented in this publication verifies our approach to isolate gravitational coupling of 100\,mg sized objects. Going forward, the achieved signal to noise ratio (SNR) of $6.04/0.06\approx 100$ already allows measuring gravity exerted by 1\,mg sized objects with $SNR\approx1$ at a measurement duration of $\approx150$\,h.
Note that our current measurement runs undergo diurnal variations in the noise floor, to which the thermal noise of our $Q=4.9$ oscillator assembly poses a hard limit. In order to make better use of our measurement time (${SNR}\propto T_{meas}$) we have to improve our understanding and the reduction of coupling of daytime environmental noise. Furthermore, mechanical 
quality values in excess of $Q=2\times10^4$ have been observed even for this particular pendulum. As the off-resonant thermal noise amplitude decreases with $1/\sqrt{Q}$, this provides a factor $\sqrt{20000/5}\approx65$ improvement over the current sensitivity.
        
Combined, this shows the potential of our approach for sensing the gravitational field of a Planck mass sized object ($m_P=22\,\mu\textrm{g}$).

\section{Calibration}
\subsection*{Optical lever readout}
The angular position of the torsion balance is read out by means of an optical lever using $1.8\,\mathrm{mW}$ of \textit{Innolight Mephisto} laser light at an angle of incidence of $\approx50\,\textrm{mrad}$. The signal is obtained as differential photo voltage of an off the shelf \textit{Thorlabs QPD80A} quadrant photo detector (QPD), which provides high dynamic range ($0.2-0.5\,\mathrm{V}_x/\mathrm{V}_{sum}$) and low noise ($\approx2\times10^{-9}\,\textrm{V}_x/\textrm{V}_{sum}/\sqrt\textrm{Hz}$) in its approximately linear region of $0.5\,\mathrm{mrad}$. The data is sampled by a Picoscope 4824 at $250$ and $286\,\mathrm{Samples/s}$ (anti-aliasing filter at $132\,\mathrm{Hz}$) and digitally downsampled by a factor 10. The ADC provides 12\,bit resolution ($\pm10\,\mathrm{V}$ full scale $\rightarrow4.9\,\mathrm{mV}$ resolution). Although we are oversampling the signal already, the high frequency quantization noise was shown to be improved when using even higher data rates. Reducing the full range would also help, yet we require it to cover large low frequency amplitudes, in particular during noisy times of data acquisition.
        
For future experiments, the actual (not quantization noise limited) readout sensitivity can be improved by using interferometric methods, although at the cost of limited absolute range. Systematic calibration errors can be significantly reduced by exchanging the current read-out with a perfectly linear, self-calibrating auto collimator at the expense of increased complexity and limited oversampling capability\cite{Turner2011picoradian}.

\subsection*{Position calibration}
Position calibration of the test mass motion was done by measuring the pendulum yaw deflection both via the optical lever and, independently, via video tracking. For the latter measurement, the torsion pendulum was recorded from the top with a resolution of $24\,\mathrm{\mu m/px}$ with both gold spheres being pattern tracked to obtain the deflection (compare Section \ref{sec:distance}). Each method contributes their own source of error: for small amplitudes, the video tracked deflection angle is relatively noise, while at large deflections the optical transfer function of the QPD readout becomes nonlinear. For our calibration, the pendulum was excited to an amplitude of $\approx 125\,\mathrm{\mu m}$, largely exceeding the linear detection range.

\begin{figure}[htb]
\centering
\includegraphics[width=1.0\linewidth]{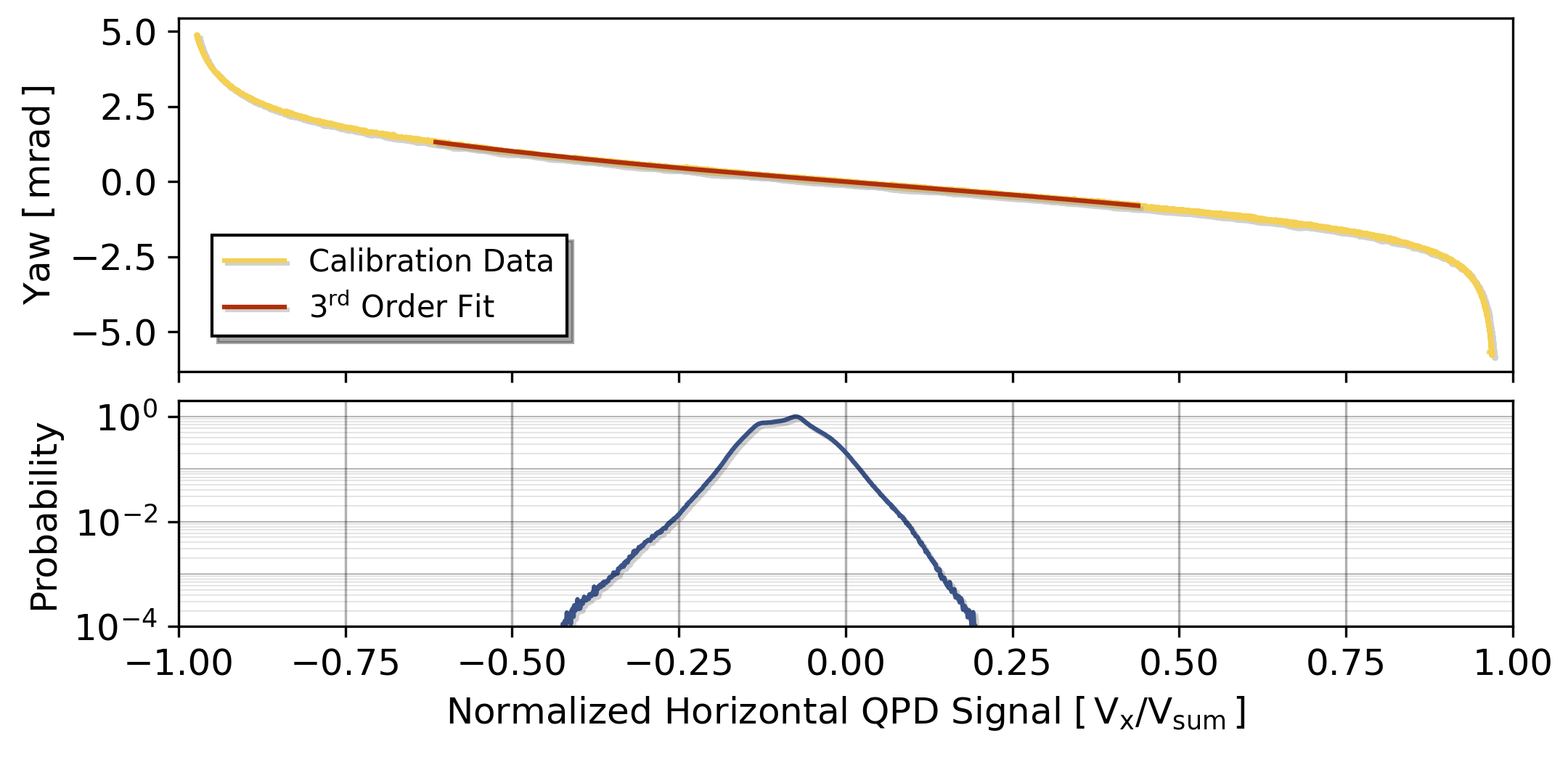}
\caption[Quadrant photodiode calibration]{The video tracked yaw angle of the torsion pendulum is compared to the QPD readout (horizontal difference signal normalized by the total sum signal to reduce laser intensity noise coupling). A third order polynomial fit to this data provides our QPD signal calibration. The occurrence probability accumulated over all our data runs shows the region in which a calibration is required, i.e.\ $V_x/V_{sum}\in[-0.4,0.2]$.}
\label{fig:PDcalibration}
\end{figure}

In the ideal case the rotation angle and the normalized QPD signal are related via the Gauss error function. In our case, clipping and other effects deteriorate this relation at large amplitudes. Since our experiment is confined to small deflections we only require a good calibration in the central QPD range. Therefore, a $3^\textrm{rd}$ order polynomial was fitted to the data surrounding $V_x/V_{sum}\approx0.09$ to serve as yaw angle calibration for all our acquired QPD data. 


\subsection*{Force calibration and force-noise estimate}\label{sec:forcecalib}
To infer the force exerted on the test mass we deconvolve the test mass position time series with the inverse of the previously deduced mechanical susceptibility. A compensation low-pass has been applied at $1\,\mathrm{Hz}$ to prevent instabilities from infinite response around the Nyquist frequency or above.



A major feature of off-resonant detection is in our case the flatness of the force noise floor around the signal. This allows to obtain a live-estimate of the current noise conditions without additional sensors. Known narrow band signals such as $f_{mod}$ and harmonics are removed from the inferred force signal and the instantaneous amplitude in the $5-75\,\mathrm{mHz}$ band is obtained by the absolute value of the signal plus its Hilbert transform. Specifically, evaluating amplitude (real part of complex signal) and change of amplitude (imaginary part) gives a continuous estimate for the noise coupling into the system. The goal of the algorithm is to identify excess excitation segments of the data that occur over a long duration, in particular to minimize the data degradation from the sometimes up to tenfold increased noise condition during daytime. We are therefore low-pass filtering (moving average, $1/f_{mod}\approx79\,\mathrm{s}$) the instantaneous amplitude and use the result for inverse variance weighting of the measurement data. Compared to completely vetoing data during bad conditions this allows to make use of the full measurement time under these extremely non-stationary conditions. Employing this method we could optimally combine a total of 350 hours of non-stationary measurement time which are worth $\approx150$ hours of higher quality data as presented in Figure \ref{fig:spectrum} and 3 into the single evaluation shown in Figure \ref{fig:G}.

\subsection*{Non-stationary environmental conditions}
\begin{figure}[htb]
\centering
\includegraphics[width=89mm]{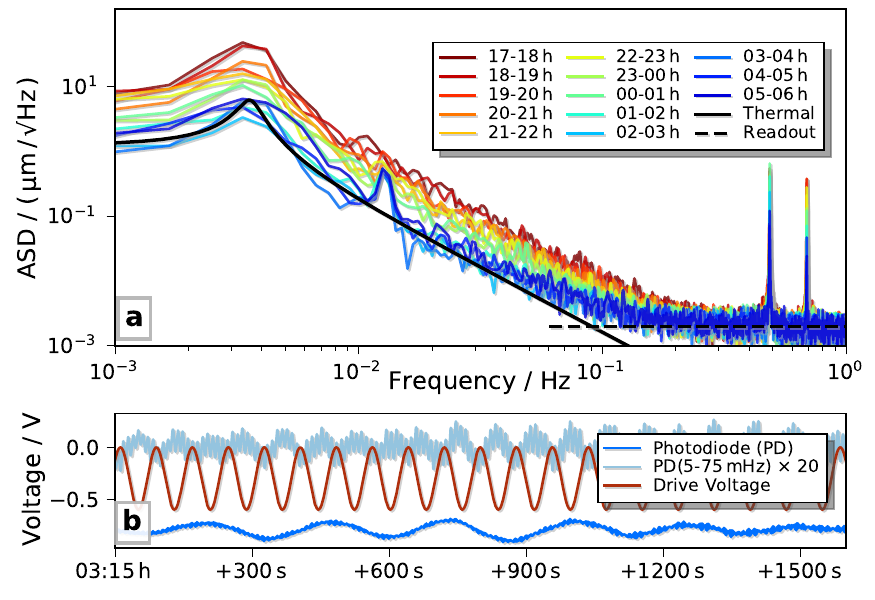}
\caption[One hour displacement spectra]{(a) The displacement spectral density estimate of representative data taken between 26\textsuperscript{th} and 27\textsuperscript{th} of December shows stationary readout noise at high frequencies and non-stationary white force noise at low frequencies. The level of the latter varies up to tenfold, consistently being the lowest at $0-5$\,a.m., in particular in the night after public holidays or Sundays / before normal workday. During such a quiet period the thermal noise of the pendulum is reached.
(b) Time series data spanning half an hour during the $3-4$\,a.m. segment. The data trace is bandpass filtered in the frequency range from $5-75\,\mathrm{mHz}$ with a 6\textsuperscript{th} order butterworth filter and magnified by a factor of 20.}
\label{fig:PSD(t)}
\end{figure}

Each data chunk contains up to 13:53\,h of data ($5\times10^4\,\mathrm{s}$). In order to display the time evolution of our noise we perform a spectral density estimate on 1\,h segments using Welch's method with 20\,min long segments and 50\,\% overlap (Figure \ref{fig:PSD(t)})\footnote{Note that the spectral resolution is inverse proportional to the integration time and therefore worse for the segmented spectra than in \ref{fig:spectrum}. Also the gravitational signal is smaller for these short segments as less signal is integrated. If we chose the amplitude spectrum representation instead, our signal would remain at the same height but the noise floor would rise as less noise is averaged out.}. 
The low frequency noise (below $\approx 100\,\mathrm{mHz}$) is dominated by actual motion of our torsion pendulum. While we reach the theoretical thermal noise limit of the damped harmonic oscillator during quiet times, we observe up to 10 times higher noise levels during other times of the day. The origin of this noise has not been fully identified yet (see also Section \ref{sec:seismic} and \ref{sec:magnetic} for more details). Still, it always exhibits the same pattern: within 1\,h segments the noise appears to be white in terms of force exerted onto the test mass. Yet its level slowly varies. The noise power is well correlated with the diurnal variability of anthropogenic noise, be it in higher frequency seismic activity (evidenced by the excitation of the 0.7\,Hz mode), or in 10\,mHz magnetic fields. Nights are better than daytimes, even with nobody being present in the lab. Weekdays present worse conditions compared to weekends and public holidays. Therefore, the Christmas 2019 time was exceptionally well suited for the low noise measurements presented in this publication.
        
At higher frequencies above $\approx 100\,\mathrm{mHz}$ the noise flattens out (except for higher order modes) at a readout noise level of $2\times 10^{-9}\,\mathrm{m}/\sqrt{\mathrm{Hz}}$. Since this noise further improves with increased sampling frequencies it is suspected to be quantization noise.



\section{Data evaluation}
\subsection*{Data handling}
All data is sampled by a Picoscope 4824 at $250$ or $286\,\mathrm{Samples/s}$ using the internal clock.
The ADC provides 12\,bit resolution ($\pm10\,\mathrm{V}$ full scale $\rightarrow4.9\,\mathrm{mV}$ resolution).
A typical data trace spans just $5\times10^4\,$s so just short of 14 hours. 
An analog anti-aliasing filter at $132\,\mathrm{Hz}$ is applied to the horizontal difference of the photodiode quandrants to get rid of high 50\,Hz harmonics (room light) and other narrow band large amplitude signals.
The QPD signals are digitally de-rotated by $\approx20$\,rad which maximizes the sensitivity to the torsion motion while minimizing the impact of pitch motion (vertical spot motion).
The calibration procedure described earlier provides a signal in terms of test mass displacement.
From this, the force signal is deduced as described below.
At this point the momentary noise estimation signal is deduced processed alongside with the other channels.
Only thereafter, the signals are digitally down-sampled by a factor 10 (a 10\textsuperscript{th} order Chebyshev Type 1 filter at $0.8f_{s,new}$ with the response at $f_{mod}$ normalized to unity is applied before the decimation) to reduce the amount of data. The initial oversampling reduces digitization noise (optimally by a factor $\sqrt{10}$).
Finally, the signals are cropped, i.e.\ usually the first 5\,\% are being removed. Besides eliminating general edge-effects of any filters, this mainly reduces the ring-in effect of the force estimation.

\subsection*{Force inference}
In general, a damped harmonic oscillator can be represented by a 2\,nd order low pass filter. The mechanical susceptibility $\chi(\omega)$ is defined as the transfer function from externally applied force $\tilde{F}_{ext}(\omega)$ to induced displacement $\tilde{x}(\omega)$.
\begin{equation}
    \chi(\omega)=\frac{\tilde{x}(\omega)}{\tilde{F}_{ext}(\omega)} = \frac{1}{m_{eff}}\left(-\omega^2-i\gamma\omega+\omega_0^2\right)^{-1}
\end{equation}
At a given frequency $\omega$, $\chi$ is defined by the (viscous) damping rate $\gamma=\omega_0/Q$, the (un-damped) resonance frequency $\omega_0$ and the effective mass $m_{eff}$. We map this to a zero pole gain (zpk) representation
\begin{equation}
    \chi\stackrel{!}{=}k\,(s-p_0)(s-p_0^\ast)
\end{equation}
in Laplace domain ($s=\sigma+i\omega$, choosing $\sigma=0$) by means of a pair of complex conjugate poles ($p_0,p_0^\ast$) and a DC-response $k$ (which contains the unit conversion). We find
\begin{equation}
    k=\frac{1}{m_{eff}}\\
    p_0=\gamma/2\pm\sqrt{w_0^2-\left(\gamma/2\right)}\quad.
\end{equation}
As long as the zpk representation of a system is known, then the time domain response $x(t)$ can be inferred from the driving force $F_{ext}(t)$ without solving the equation of motion by the following recipe e.g.\ using \textit{MATLAB}'s built-in functions:
\begin{enumerate}
    \item convert the zpk model to its transfer function form by \texttt{[den,num] = zp2tf(z,p,k)} (appropriate k must be chosen to take into account the re-normalization by placing poles/zeros!)
    \item convert continuous time model to filter coefficients of discrete time model sampled at frequency $f_s$ by \\\texttt{[bd,ad] = bilinear(den,num,fs)}
    \item apply the filter by \texttt{x = filt(bd,ad,F)}
\end{enumerate}
While steps 1\&2 are more or less technical re-definitions, step 3 convolves the force time series with the impulse response in a very resource friendly manner.

We use this method in reverse by inverting the mechanical susceptibility, which is equivalent to exchanging poles and zeros $p_0\rightarrow z_0^{\prime}$, $z_0\rightarrow p_0^{\prime}$ and $k\rightarrow 1/k^{\prime}$ to infer the force $F_{ext}$ required to produce the measured displacement $x(t)$. In order to avoid an un-physical (infinite) response at or above the Nyquist frequency $f_N=f_s/2$ from the two zeros, we have to limit the response by a pair of compensation poles. By allowing them to be complex as well, their phase loss around the detection band can be limited. In step 2, we apply pre-warping at 100\,mHz to reduce malicious effects onto the transfer function. The compensation poles were chosen to lie at $w_{comp}=2\pi\times1\,$Hz with $Q_{comp}=3$. This choice was verified not to influence the measured gravitational signal significantly (compensation pole in Table II).

\subsection*{Data visualization}
The spectrum estimate shown in Figure \ref{fig:spectrum} was produced by Welch's method (\texttt{pwelch}). Each time series is divided into five blocks with 50\,\% overlap and tapered by a Hanning window. The spectrum estimates of the blocks are then averaged. Hence, the noise floor in the figure is elevated but smoothed out compared to performing a single spectral estimate on the whole trace at once.
The residuals shown in the plot are a spectral estimate of the difference of the measured force signal (red trace) and the expected/nominal gravity signal. By performing the subtraction before the spectral (power) estimate, these residuals become sensitive to the relative phase of the two signals.

As our goal is to quantify a (nonlinear, but stationary) relation between source-test mass separation and exerted force, plotting one against the other comes with the implicit assumption of equal significance of each data point. As our momentary noise estimation channel, however, provides us with means of judging the significance (in therms of signal to noise ratio) of each point, we chose to weight the data points in Figure \ref{fig:bonfire} with this significance to obtain the most general visual representation of the force versus separation relation. The hidden assumption here is that the momentary noise level estimation procedure does not introduce systematics into the quantitative data evaluation (see next step). The authors tried to verify this by checking that there is no quantitative dependence of the results from noise estimation filter parameters.

Only after this most general data representation, the relation is subjected to assumptions by using Newton's law to quantify the relation. Hence, the width of the signal band is determined by all noise within the bandwidth of evaluation ($5-75\,$mHz bandpass, see above) and not only at the signal frequency or its harmonics. While subjecting data observed in dedicated high precision experiments to the rigid corset of assumptions (fitting parameters) is a well established method, we believe the least assumptions data representation approach of Figure \ref{fig:bonfire} to be extremely valuable in a yet unexplored regime of gravitational sensing.

\subsection*{Data evaluation}
In order to quantify the measured gravitational force we assume the validity of Newton's law and treat the masses as point masses concentrated at their respective centers of mass. Strictly speaking, this assumption is valid only for perfectly rotationally symmetric masses. Possible deviations due to bumpyness are considered as systematic errors. We allow the gravitational constant as fitting parameter resembling the strength of $1/x_s^2$ coupling. We have to further allow for a distance independent force resembling the measured force averaged over one signal period as the experiment was not designed to measure the DC-attraction, i.e.\ the force with relation to the source mass being infinitely far away.\\
The statistical uncertainty of data can be inferred by resampling it. While often bootstrapping is used, it turns out that without further pre-conditioning high frequency temporal correlations such as high frequency vibrational modes spoil the results as the data points cannot be regarded as sufficiently independent. Subsampling, in contrast, holds under weaker assumptions and appears to work for our data as short term correlations remain within the same segment.\\
Each time trace is cut into $N$ equally long segments. For each segment, the coupling strength $G_i$ is fitted with inverse variance weight factors described in the previous section non-stationary environmental conditions. Furthermore, a common force offset $F_0$ is fitted to all segments (allowing each segment to have its own DC-force $F_{0,i}$ leads to similar results, but is less physical as this force shouldn't drift). The weighted mean
$G=\sum_i G_i/\sigma_i^2$, where $1/\sigma_i^2$ are the normalized inverse variance weights for the segment, gives the same result as a weighted fit to the whole trace. From the (inverse variance weighted) standard deviation $std_w(Gi)$ we infer the statistical uncertainty of each $Gi$. The uncertainty for the whole trace is then inferred by means of $\mathrm{std}(G)=\mathrm{std}_w(G_i)/\sqrt{N_{eff}}$ with an effective number $N_{eff}$ of segments with independent noise contributing to the weighted standard deviation. The number $N$ of segments per trace is varied between 11 and 30. For small $N$, drifts within the segments and non-stationarity of noise within the segments may become significant, while for large $N$ the exact resonance frequency and quality factor of the torsion resonance gain importance as the segment length approaches a torsion period. To exclude such effects, for each trace the results were verified not to depend on $N$ in order not to introduce systematic errors by its exact choice and the fit parameters were averaged over the full range of segmentations. 

This evaluation gives the independent estimate of the coupling constant and estimate for statistical uncertainty for each trace presented in Figure \ref{fig:G}. All deduced coupling constants agree well within their statistical error budgets. A coupling constant with even lower statistical uncertainty is obtained by averaging the coupling constants of all traces while using the respective deduced statistical uncertainties as inverse variance weights. The power in the time resolved representation of Figure \ref{fig:G} is that it shows that there are no major (monotonous) time-dependent systematics such as adsorption, continuous electrical charging etc.

\section{Vacuum}
The experiment has been conducted in high vacuum ($10^{-6}-10^{-7}\,\mathrm{hPa}$) to reduce excitation from residual gas molecules and direct momentum transfer. 
The vacuum chamber is pumped by a turbo molecular pump (\textit{Pfeiffer TMU 071 P}) running at $25\,\mathrm{Hz}$ during measurement runs. These narrow band vibrations have not been observed to produce sensitivity degradation of the gravitational signal.
The pressure could not be monitored permanently during measurement runs as the Pirani pressure gauge (\textit{Pfeiffer PKR 251}) had to be dismantled from the chamber. It had been observed to constantly charge the torsion balance when in use and its relatively strong magnetic field was suspected to cause magnetic interaction between the gold masses.

Impacts of residual gas molecules cause a Brownian force noise that is derived as an additional damping rate \cite{Schmoele2017}: 
\begin{equation}
    \gamma_{wall}=\frac{P}{m v_{air}} r^2 \frac{\pi^{3/2}r^2}{\sqrt{2}d^2ln(r^2/d^2+1)}
\end{equation}
(P: pressure; $v_{air}=k_B T/m_{air}$: average velocity of 'air molecules' of mass $m_{air}$; d: separation between the surface of the cylinder and the EM-shield; r: cylinder radius; m: cylinder mass; $k_B$: Boltzmann constant; T: Temperature).
The additional force noise damping due to the proximity to the EM-shield is taken into account. Moreover, it assumes the worst case scenario of a cylindrical test mass, which would trap considerably more gas molecules compared to a sphere of similar size close to the wall.
Using our experimental parameters (P=$6\times10^{-7}\,$mBar, $m_{air}=4.8\times10^{-26}\,$kg, $r=1\,$mm, $m=91$mg, $d=150\,\mu$m, $T=300\,$K, ) we obtain an additional damping of $\gamma_{wall}\approx 5.1\times 10^{-8}$, which is much smaller than the natural damping rate $\gamma_{m_t}=\omega_0/Q\approx 5.0\times 10^{-3}$ of our test mass oscillator.
Looking ahead, even for a projected experiment in the Planck-mass regime, which would require a mechanical quality $Q>20,000$, the effect of force noise from gas collisions would be at the 5\% level in this pressure regime - and can be further reduced by pushing vacuum levels into the UHV.

\section{Vibration isolation}
The vacuum chamber is placed  on an optical table with deflated air springs. While inflating the springs did attenuate high frequency noise as expected\cite{Newport}, the pendulum yaw mode in contrast was observed to get heavily excited. We suspect a relation to table tilt, which we have independently observed to couple strongly into the experiment ($3\,\mathrm{rad}_{yaw}/\mathrm{rad}_{tilt}$). In the inflated state air density fluctuations act differently onto the air springs and thereby can cause excessive table tilt at low frequencies, which could explain the excessive yaw excitation. Operating our experiment with inflated air springs will require an active tilt stabilization system as described in Ref.~\cite{lewandowski2020active}
        
High frequency vibrations are known to couple non-linearly into lower frequency bands, where the measurement takes place\cite{Shimoda2019}. To decouple the pendulum from such vibrations, for example as generated by the turbo pump, the $420\,\mathrm{g}$ pendulum support structure is resting on soft vacuum compatible rubber feet (\textit{Viton\textsuperscript{\textregistered}}). 

\section{Source mass position drive}
\begin{figure}[htb]
\centering
\includegraphics[width=45mm]{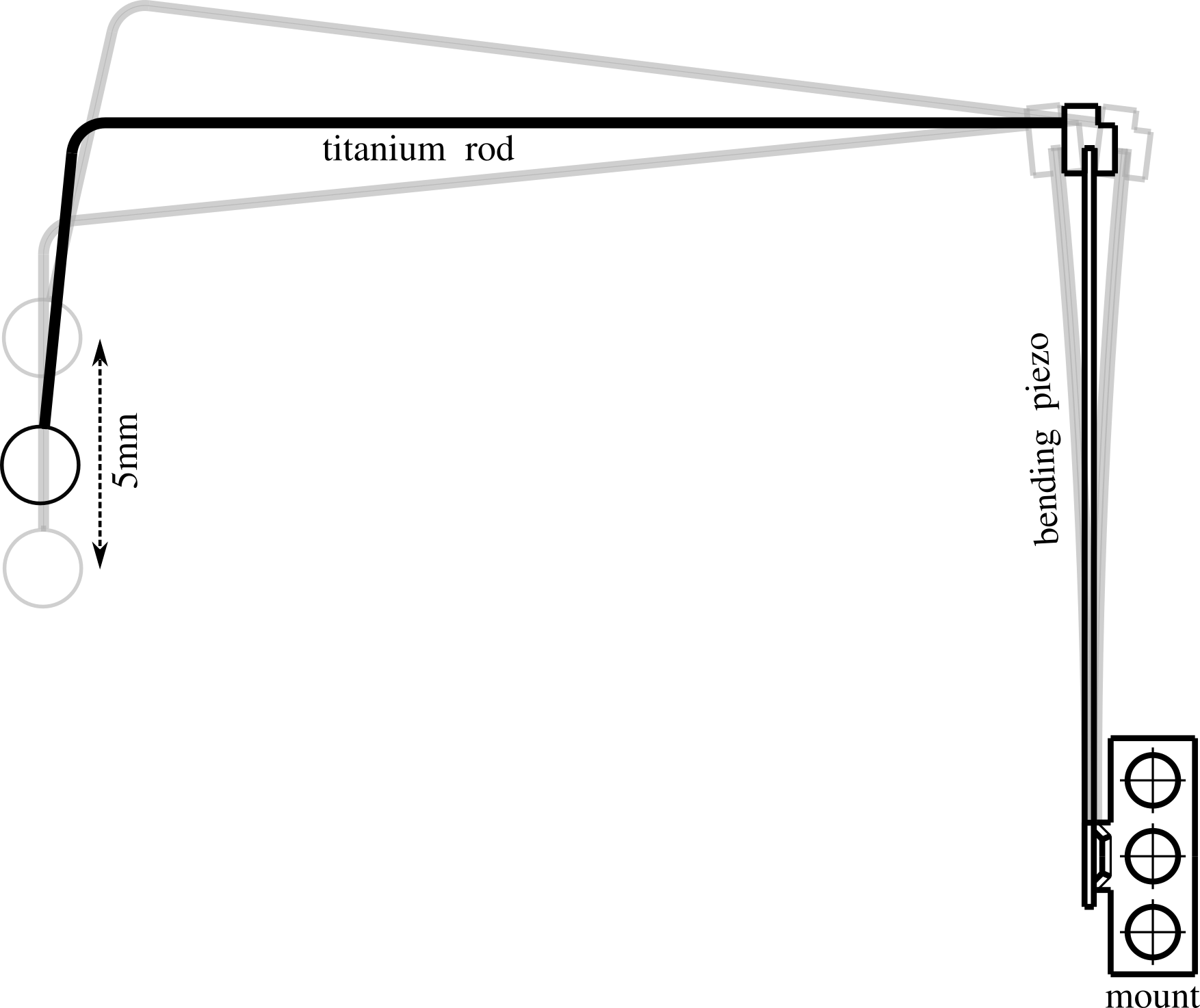}
\caption[Schematic of drive mechanism]{The source mass drive consisting of a bending piezo and a titanium rod is able to modulate the source mass position by more than $5\,\mathrm{mm}$ peak-peak at frequencies as high as 1\,Hz. The geometry of the titanium rod amplifies and translates the piezo deformation into an approximately linear motion.}
\label{fig:drive}
\end{figure}

The position of the source mass gold sphere is modulated by a long range bending piezo (\textit{PI PL140.10}, PICMA\textsuperscript{\textregistered} Bender). Applying a voltage of $0-60\,\mathrm{V}$ bends the  piezo and provides on the order of $1\,\mathrm{mm}$ travel range at the piezo tip, which is connected to a $0.31\,$mm diameter non-magnetic titanium rod carrying the source mass (Figure \ref{fig:drive}).

The rod was inserted only halfway into the source mass so that its contact potential facing towards the test mass was effectively shielded by the surrounding gold.  
        
Upon actuation, the piezo bending is translated to a mostly linear motion of the source mass of up to $5\,\mathrm{mm}$ and a rotation around its center, which can be neglected. The position of the source mass is modulated harmonically in order to avoid DC-noise of the torsion pendulum such as slow thermal drifts. 


\section{Distance measurement}
\label{sec:distance}

A major source of uncertainty for the measurement of the gravitational coupling $G$ is the absolute separation between source mass and test mass.
We measure the distance with a DSLR camera (\textit{Canon EOS 60D}) that is mounted on top of the vacuum chamber and is looking down vertically onto the experiment. 
This means that only a 2D projection of the assembly can be monitored.
Videos were taken with a $60\,\mathrm{mm}$ macro lens and extension tubes. 
The positions of the masses were tracked using a template matching algorithm based on the open-source image and video processing library \textit{OpenCV}.
            
At the beginning of each distance measurement, templates of the masses are defined in the first frame of a video. Their positions are tracked in subsequent frames by integrating the squared pixel value distance over the dimensions of the template
\begin{equation}
    R(x, y) = \sum_{i\in \{r,g,b\}} \left(\sum_{x', y'} \left( T_i(x', y') - I_i(x + x', y + y') \right)^2\right).
\end{equation}
Here, $T_i(x, y)$ designates the 8-bit pixel value of the colour channel $i$ of the template image, and $I_i(x, y)$ the same for the video frame to be evaluated.
The minimum $\min(R) = R(\tilde x, \tilde y)$ then corresponds to the position of the best match $(\tilde x, \tilde y)$ for the template.
This method is limited to pixel accuracy with $1\,\mathrm{px} \approx 24\,\mathrm{\mu m}$. Therefore, for each dimension a Gaussian fit over the five neighboring pixels centered at the best match position is performed to improve on the tracking resolution.

In order to limit the effect of changes in lab lighting as well as the influence of reflexes on the source mass surface, only the outer perimeter of the masses is tracked. We verify the tracking quality by visual inspection of the residuals, i.e.\ by looking at a video with each frame realigned to the tracked position similar as in a digitally stabilized video. The resolution and accuracy of the tracking is limited by the (manual) determination of the edges of the masses in the template, the low contrast inside the vacuum chamber due to its gold plated interior and the resolution of the DSLR and is taken to be $20\,\mathrm{\mu m}$.

We measured the separation of source and test mass via video at the beginning and the end of each measurement run.

Furthermore, the source to test mass separation is not only defined by the momentary source mass position, but also by residual test mass motion. In particular, the center of mass motion of the pendulum varies the distance by more than $100\,\mathrm{\mu m}$ under bad conditions. Therefore, the gravitational force is averaged over these distances. The separation modulation changing from this test mass motion cannot be extracted from the optical lever signal to be calibrated out. In future continuous video tracking will provide this information.

\subsection*{Piezo calibration}
The piezo is operated without feedback. As the position could not be monitored continuously, the piezo's position response to the applied voltage needs to be characterized and modeled.
The response of the piezo can be modeled as an ideal mechanical transducer combined with an additional low-pass filter\cite{creepyPI}.
Fundamentally, this reflects electron mobility / diffusion into the piezo material. 
As a consequence, the step response is delayed and the drive shows a slow creep towards its final position as well as hysteresis when driven cyclically.
The creep (knowledge of the exact position) is of particular interest for the charge measurements while the hysteresis (knowledge of the exact amplitude and phase) is important for the gravity measurements, as it results in the source mass position lagging behind the applied voltage signal and reduces the amplitude.
The stationary creep (the step response of the piezo) depends on the size of the applied voltage step and the starting position (the voltage levels applied to the piezo before and after) and can reach up to $30\,\mathrm{\mu m}$ over a period of 15 minutes.
As the creep depends on parameters we have little control over, the creep has been fitted for every measurement separately rather than using global piezo parameters.
The observed time lag depends on the drive frequency, but does not show any dependence on the modulation amplitude.
For a modulation frequency of $12.7\,\mathrm{mHz}$ we find a time lag of $(509\pm3)\,\mathrm{ms}$ by correlating the drive voltage with the source mass position as measured by video tracking.
        
We also observe a drift over time in both the mean position and the modulation amplitude of the source mass drive. A possible reason is continuous charging of the piezo material, similar to the observed creep for a step response in quasi-static operation even in dynamic operation.
In contrast to the stationary case where creep induced piezo deflection reached a final state after about an hour, under cyclic operation the mean deflection and amplitude continued to drift over a measurement period of 100 hours.
For a peak-to-peak modulation of about $2\,\mathrm{mm}$, the amplitude reduced by $13\,\mathrm{\mu m}$,  while the mean position drifted by $55\,\mathrm{\mu m}$.
Although this long term drift resembles the stationary creep as described in\cite{creepyPI} with a much longer time constant, we cannot fit the corresponding function with only the start and end videos.
The drift of the mean position as well as the modulation depth, therefore, is taken to be linear over the period of a measurement run.

Additionally, the source mass position modulation exhibits higher harmonics contributions. We measure the relative amplitudes and phases of these higher harmonics with respect to the fundamental mode in the start and end videos.
We infer the actual position modulation for the gravitational measurement runs from the drive signal by taking the time lag, the amplitude change and mean position drift and the higher harmonics into account.
For example, the amplitude of the 2\textsuperscript{nd} harmonic of the drive position has been fitted to and found to be below $0.3\,\%$ of the fundamental amplitude. Remaining higher order contributions are estimated to be below $<0.1\,\%$ and are therefore irrelevant for our experiment.


\section{Charge measurements}\label{sec:charges}

We are discharging the test mass to less than $8\times10^4$ elementary charges using a nitrogen gas discharge and ion diffusion as developed in Ref.~\cite{discharging}.
The nitrogen gas is channeled over a set of high-voltage AC tips. After passing through an aperture, the nitrogen ions neutralize the surfaces of the vacuum chamber by diffusion.
After discharging, the unshielded electrostatic interaction is still approximately three times stronger than gravity (see Figure \ref{fig:shieldcomparison}).
Note that the interaction is enhanced over a pure Coulomb interaction by induced surface charges of the source and test mass.

\begin{figure}[htb]
\centering
\adjustbox{trim={0\width} {0\height} {0\width} {0.05\height},clip}%
{\includegraphics[width=0.9\linewidth]{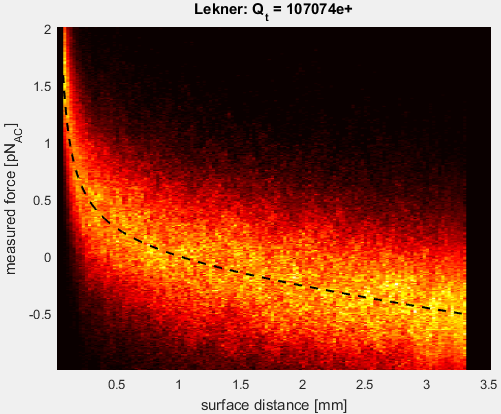}}
\caption[Probability distribution of F(d) without EM shield]{Without electrostatic shielding, the probability density distribution of inferred force versus source- test mass separation requires three constituents to describe: Newtonian gravity (see Figure \ref{fig:bonfire}), electrostatic interaction between charged ($\approx 8\times10^4\,e^+$) test- and grounded source mass\cite{leknertheo} and an unexplained force proportional to the separation.}
\label{fig:Lekner-bonfire}
\end{figure}

Because of that, even if the source mass was perfectly grounded, a charged test mass induces charge separation resulting in parasitic attraction.
While most torsion experiments striving to measure small forces conductively coat their suspension filament and thereby sacrifice mechanical quality (statistical error) for charge control (systematic error), we explicitly chose not to do so. One reason is that when down-scaling the torsion balance further the proportion of lossy coating (surface) grows with respect to fiber material (volume). The other is the long term prospect to contact-less levitate the test mass. This will not allow for any sort of electrical contacting at all.

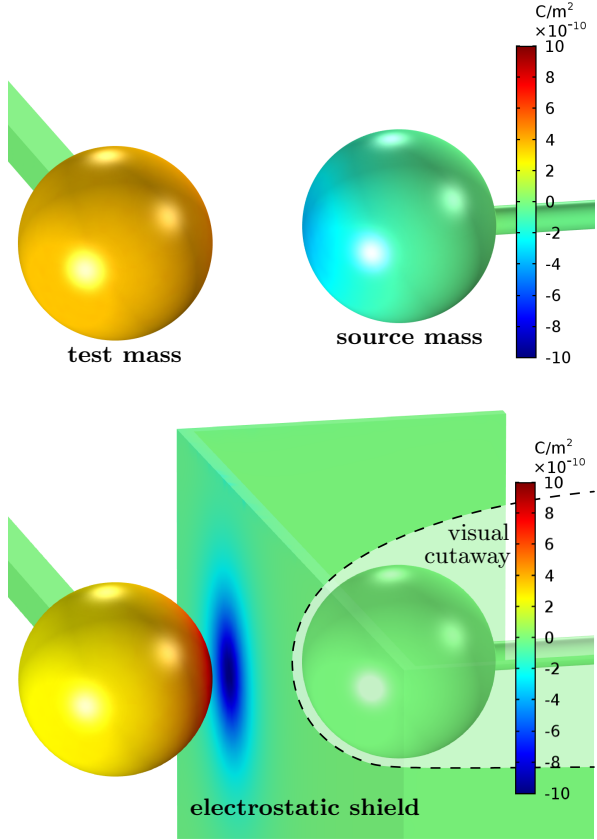
\begin{figure}[htb]
  \centering
  \begin{overpic}[width=0.9\linewidth]{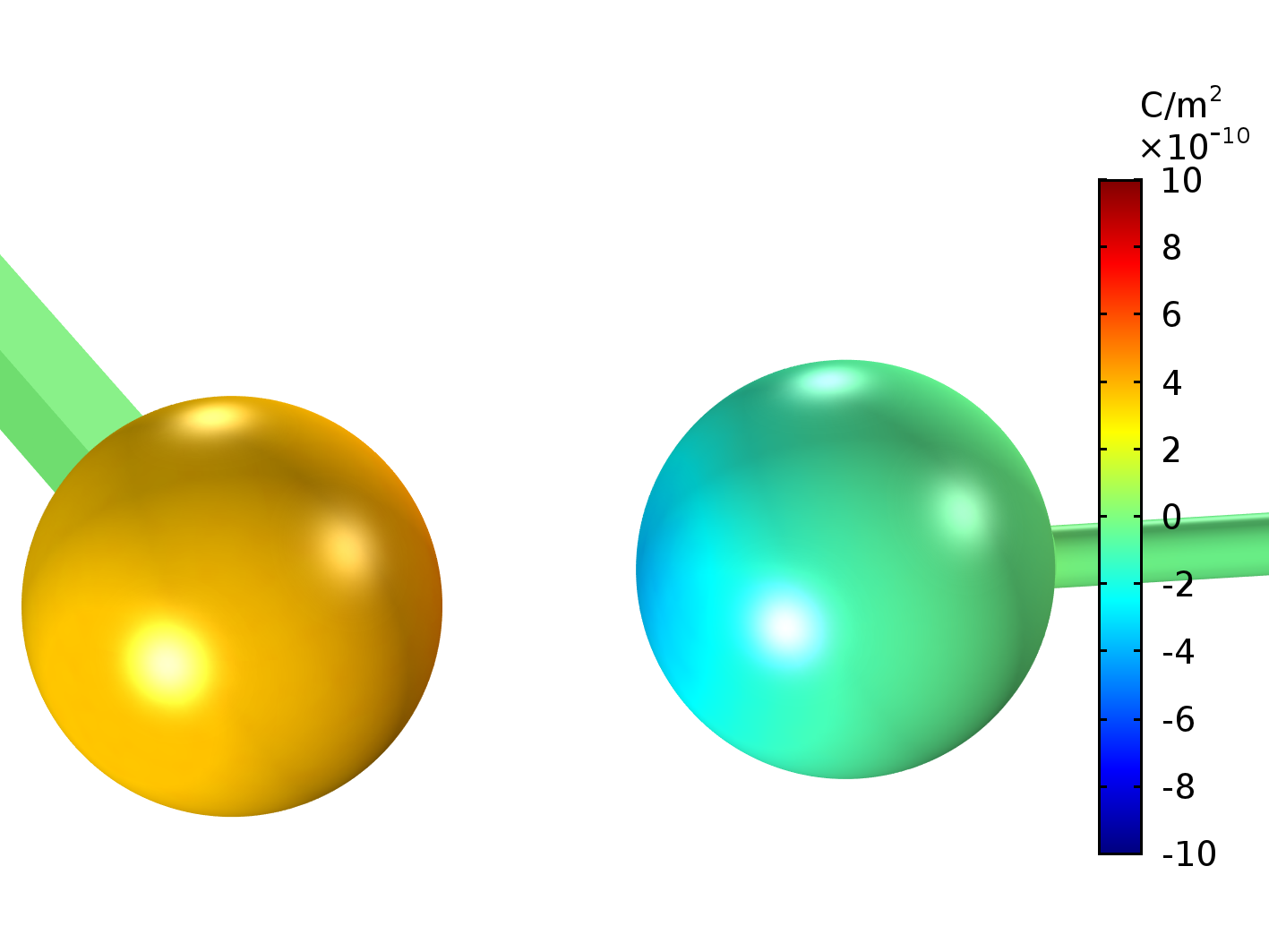}
    \put (10,7) {\textbf{test mass}}
    \put (56,10) {\textbf{source mass}}
  \end{overpic}
  \centering
  \begin{overpic}[width=0.9\linewidth]{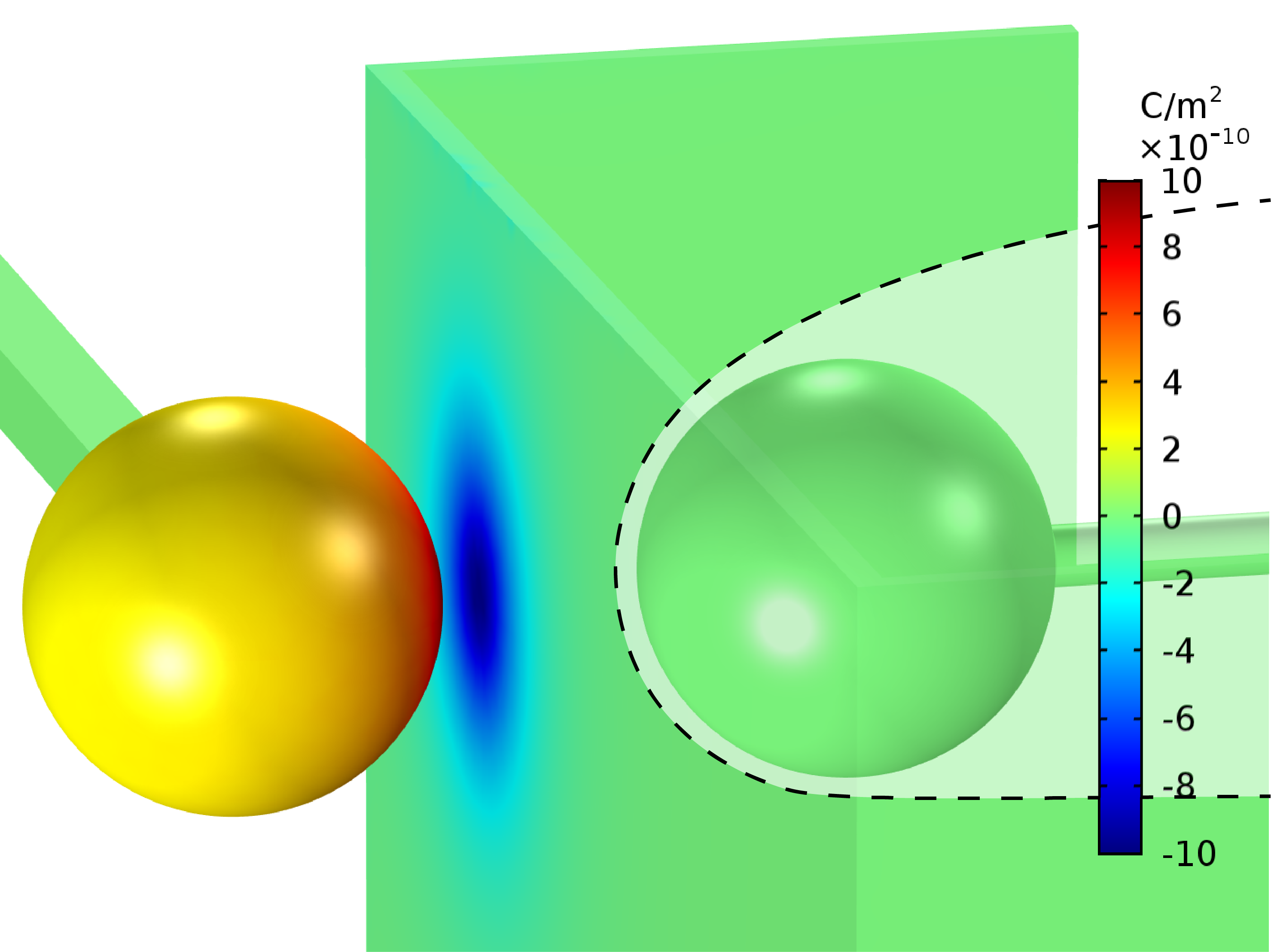}
    \put (75,51) {visual}
    \put (71,47) {cutaway}
    \put (31,4) {\textbf{electrostatic shield}}
  \end{overpic}
  \caption[FEM simulation of charge distribution]{The charge distribution induced by a $3\times10^4$\,e$^+$ charged source mass onto the grounded test mass was determined by a finite element simulation (\textit{COMSOL}, exemplarily shown for 1\,mm surface separation). With the grounded, conductive electrostatic shield in place a lot of mirror charges are induced in the unmodulated shield, resulting in a DC-force. The induced surface charge density (color coded) on the position modulated source mass is suppressed approximately by a factor 100. These charges are screened by the shield resulting in further suppression of the exerted force. The actual suppression could not be quantified due to numerical inaccuracies.}
  \label{fig:FEM}
\end{figure}

It could be shown by finite element simulations (Figure \ref{fig:FEM}), however, that electrostatic coupling between source mass position and test mass force can be suppressed well below 3\,\% of the expected gravitational signal by means of a $150\,\mathrm{\mu m}$ thick conductive Faraday shield inserted in between source and test mass (see Figure \ref{fig:shieldcomparison}). The shield comes at the added benefit of reducing other electrodynamic interactions such as short range Casimir forces as well.

\begin{figure}[htb]
\centering
\includegraphics[width=1.0\linewidth]{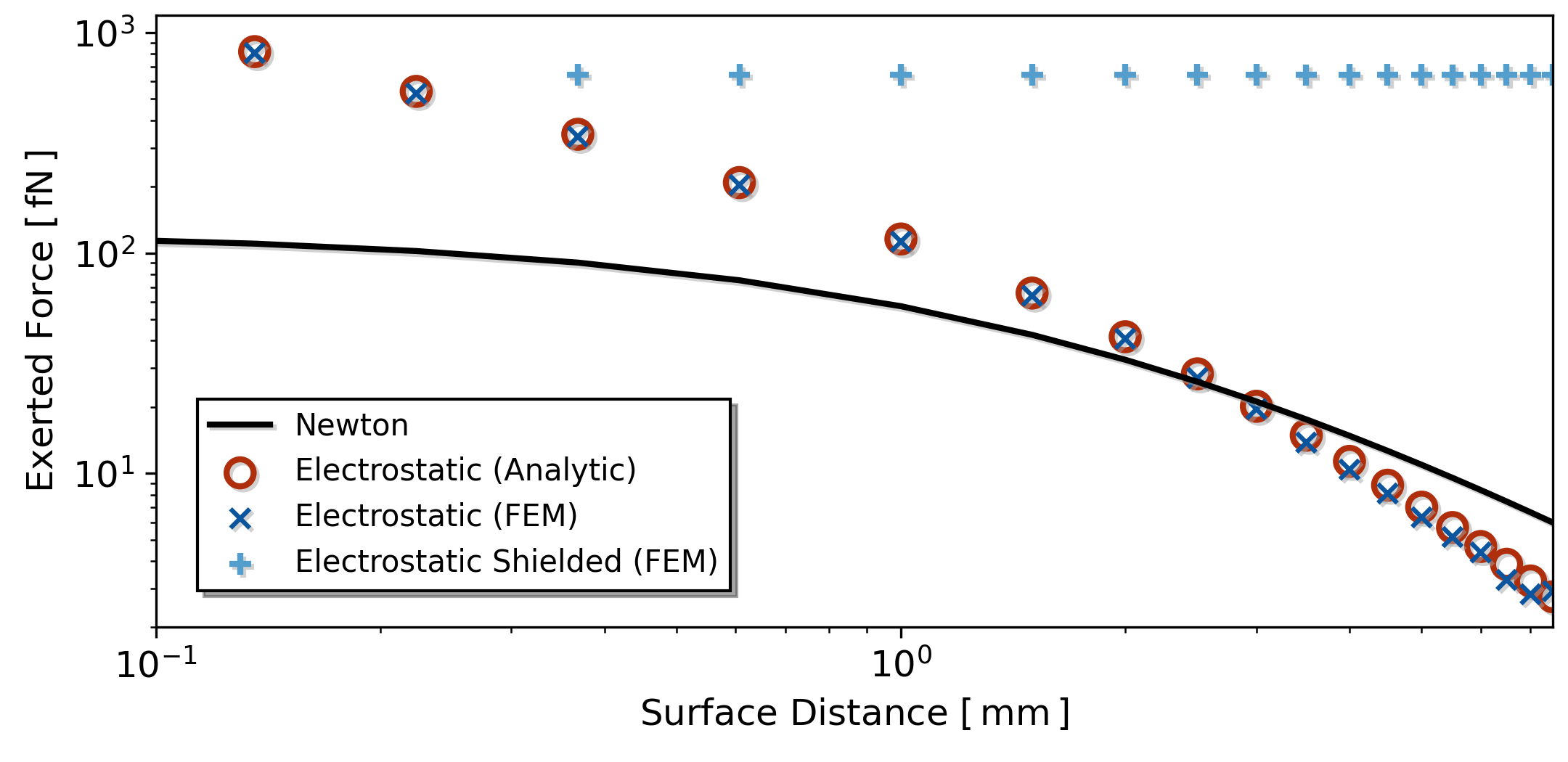}
\caption[Simulation of electrostatic coupling]{Our finite element simulation of electrostatic interaction of a charged test mass ($10^5$\,e$^+$) and a grounded source mass was validated using the analytic method described in Ref.~\cite{leknertheo}. When inserting a $150\,\mathrm{\mu m}$ thick, conductive electrostatic shield in between, the electrostatic force exerted onto the test mass is dominated by test mass mass to shield interaction, i.e.\ becomes source mass position independent. Residual fluctuations of the force are on the $3-10\,\%$ level of gravity without a clear source position dependence and are expected to stem from numerical errors.}
\label{fig:shieldcomparison}
\end{figure}

The drive piezo was housed in a Faraday shield in order to suppress coupling of the applied electric fields onto the pendulum e.g.\ via residual charges or polarization effects.
During an 8.5\,h measurement without a source mass attached but piezo modulation active, we did not observe a test mass actuation. 
While this verified the effectiveness of the shield, a full analysis of this drive piezo contribution to the systematic error of our experiment would require the same amount of effective integration time as in the gravity run, i.e.\ a measurement run would have to last for weeks under quiet conditions as well.

As long term drifts from charge diffusion on non-conductive materials were observed to degrade our low frequency sensitivity, all surfaces inside the vacuum chamber (except for optical viewports) were made from conductive materials and gold plated wherever possible to equalize surface potentials and mitigate thermo-electric potentials.

\subsection*{Charge characterization}\label{sec:rescharges}
Static charges on our test mass can be characterized by means of induced mirror charges. They can be quantified by a force versus separation measurement without electrostatic shield and with the source mass being grounded as shown in Figure \ref{fig:Lekner-bonfire}. This is effectively a macroscopic realization of a Kelvin probe force microscope, which is typically used to contact-free measure (surface) potentials. The theory for the electrostatic interaction including polarization effects of extended conductive spheres is presented in Ref.~\cite{leknertheo}. In addition to the expected signals (electrostatic, gravity) we observe an unidentified coupling which is proportional to the sphere separation. 
        
To overcome this and other experimental challenges such as unknown contact potentials we chose to isolate the effect induced by residual test mass charges. We therefore quantify them by biasing the source mass potential and observing the test mass charge vs. source mass potential interaction at a set of discrete separations.
Table \ref{tab:chargetable} summarizes the results of the intermittent charge measurements over the whole duration of the measurements.
A more detailed description of the charge determination will be presented in an upcoming publication.

In conjunction with the electrostatic shield between the masses and our finite element simulation we have thereby verified that electrostatic coupling is suppressed to below $3\,\%$ of gravity over the whole measurement period.



\begin{center}
\begin{table*}[htbp]
\centering
 \begin{tabular}{|l || @{\hskip .5em}c@{\hskip .5em}|@{\hskip .5em}c@{\hskip .5em}|@{\hskip .5em}c@{\hskip .5em}|@{\hskip .5em}c@{\hskip .5em}|@{\hskip .5em}c@{\hskip .5em}|@{\hskip .5em}c@{\hskip .5em}|} 
 \hline
  & 12/28/2019 & 12/23/2019 & 12/27/2019 & 01/03/2020 & 01/07/2020 & 01/15/2020 \\ [0.5ex] 
 \hline\hline
 surplus charges $[\,10^3\,\textrm{e}^+]$ & $34.7$ & $25.0$ & $44.4$ & $78.8$ & $38.8$ & $52.8$ \\
 \hline
 accuracy $[\,10^3\,\textrm{e}^+]$ & 0.7 & 2.6 & 9.4 & 0.4 & 3.9 & 1.1 \\ 
 \hline
\end{tabular}
\caption[chargetable]{The surplus charges on the test mass have been measured in between our measurement runs during the time periods marked as grey bands in Figure \ref{fig:G}. All measured charges stay well below $8\times10^4\,e^+$, which we take as an upper limit in our estimation of electrostatic coupling between source and test mass.}
\label{tab:chargetable}
\end{table*}
\end{center}

In conjunction with the electrostatic shield between the masses and our finite element simulation we have thereby verified that electrostatic coupling is suppressed to below $3\,\%$ of gravity over the whole measurement period.

\section{Seismic noise}\label{sec:seismic}
Even during the environmentally quiet Christmas season our fundamentally thermal noise limited sensitivity is dominated by a yet unidentified contribution with a strong diurnal variation (see Figure \ref{fig:PSD(t)}). Reduction of this noise is important to advance towards smaller detectable source masses.
        
Long period seismic ground motion in an urban environment is a major concern for extremely sensitive low frequency experiments\cite{diaz2017urban,groos2009,bourdillon2015opposite}.
For ideal harmonic motion, the rotational mode of an oscillator will not be directly excited by translational ground motion. Unfortunately, unavoidable imperfections in the construction of the pendulum cause asymmetries that couple horizontal ground motion into rotational motion\cite{shimoda2018seismic, gettings2020air}. Furthermore, down conversion of noise at higher frequencies such as cross coupling from other modes cannot be neglected\cite{shimoda2019nonlinear}.

To monitor horizontal and vertical ground displacement down to $100\,\mathrm{mHz}$, a \textit{Raspberry Shake 3D} seismometer has been installed in close proximity to the experiment. Low frequency seismic displacements at the experiment site were monitored with a \textit{RefTek Observer 60s} broadband seismometer. In addition, we have access to the broadband data of the ZAMG seismic station (\textit{Streckeisen STS2}) at Hohe Warte, Vienna, roughly $3\,\mathrm{km}$ away from the lab.

Correlating strong seismic events like the 6.7 mag earthquake in Turkey on January, 24th 2020 in the data of the local seismometer as well as the STS2 with the response of our oscillator provides a rough estimate of seismic coupling, which suggests that the diurnal low frequency noise observable in our data is not due to linear coupling of horizontal or vertical ground motion.

\begin{figure}[htb]
\centering
\includegraphics[width=1.0\linewidth]{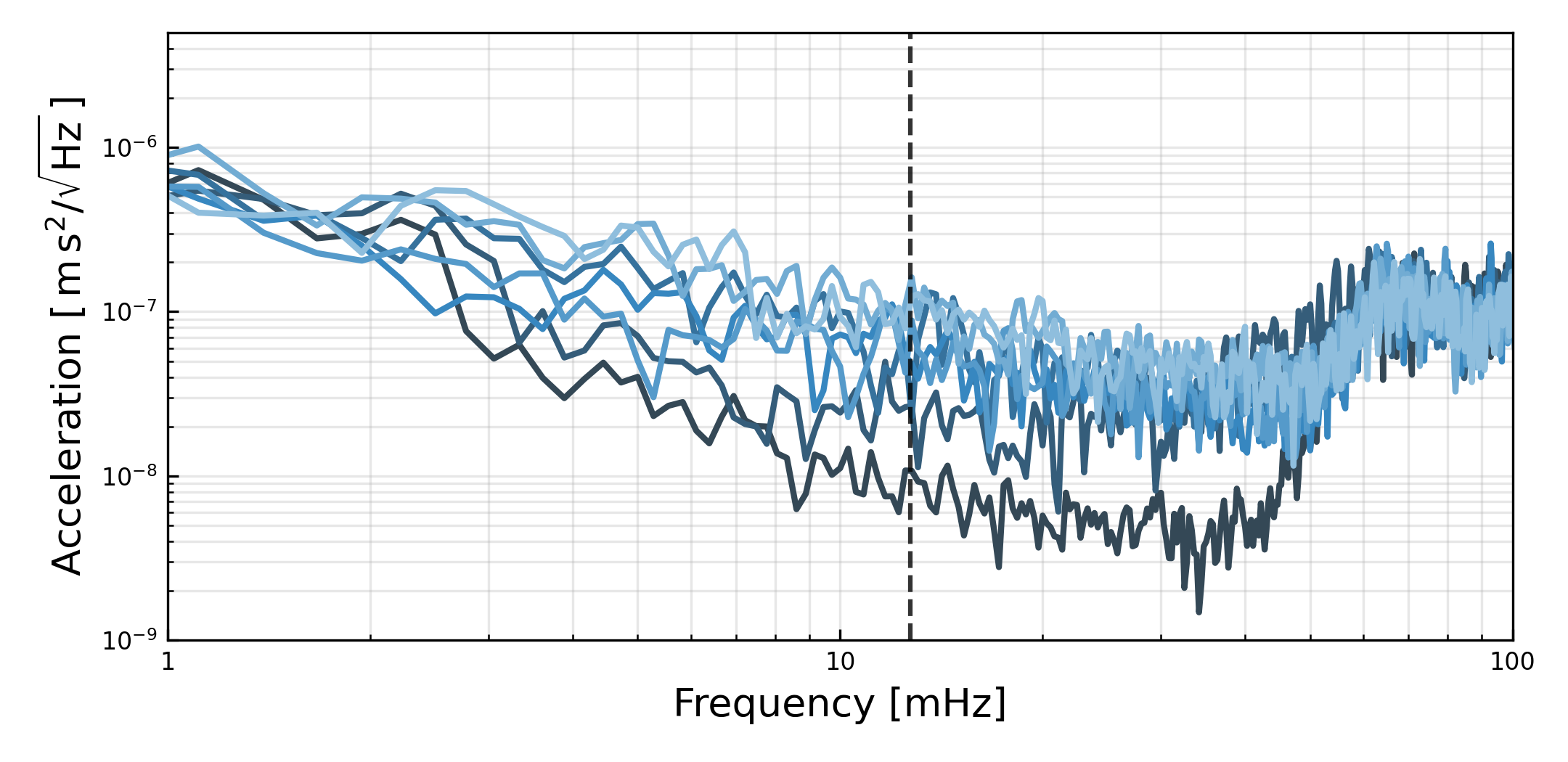}
\caption[seismicLow]{The time resolved amplitude spectral density of the horizontal ground motion in Vienna recorded by an STS2 broadband seismometer shows a strong frequency dependence of the variability. A rise of the noise floor is observed throughout the entire frequency range, starting from midnight (dark blue) to noon (light blue), but is most prominent at $30-40\,\mathrm{mHz}$. By contrast, the noise floor at the microseismic peak, at around 70\,mHz, varies only slightly. The modulation frequency of the drive mass is marked by the vertical dashed line.}
\label{fig:seismic2}
\end{figure}

This claim is supported by directly comparing the non-stationary low frequency noise in the STS2 data in Vienna (Figure \ref{fig:seismic2}) and the torsion balance (Figure \ref{fig:PSD(t)}).
There is a strong variability of the seismic noise floor in the $30\,\mathrm{mHz}$ band but a relatively steady primary microseismic peak around $70\,\mathrm{mHz}$.
The daytime dependent change in the $30\,\mathrm{mHz}$ noise floor is consistent in amplitude and time with the varying white force noise that we see in our test mass below $100\,\mathrm{mHz}$. Yet, the frequency distribution is very different. In particular,  the microseismic peak at $70\,\mathrm{mHz}$ is a very strong, stationary seismic feature within the frequency band of the experiment. Since we do not see it in the torsion pendulum data, we conclude that our rotational mode is not dominantly driven by linear coupling of the horizontal seismic noise.

\section{Mechanical coupling}
Recoil of the source mass drive is of major importance for higher modulation frequencies such as $50\,\mathrm{Hz}$\cite{Schmole2016}. Here we argue that due to its strong frequency dependence this effect can be neglected for our experiment.
A 100\,kg freely suspended optical table for example would move by 1\,nm per mm of 100\,mg source mass motion from recoil which is on the order of our observed signal. The coupling into pendulum yaw is attenuated further by the pendulum's common mode rejection, i.e. how well balanced the assembly is. Yet, already a table suspension resonance of 1\,Hz which can roughly be assumed for the air springs, attenuates this by another factor $(f_m/f_0^{table})^2\approx1/6000$. A deflated table, in contrast, typically exhibits internal resonances well beyond the 100\,Hz region yielding even much higher resistance to recoil. 

Another possible coupling at low frequencies is the cyclic deformation of the environment under the changing load distribution of the drive setup. 
Again, imagine the 1\,m sized 100\,kg optical table suspended by air springs ($\approx4\times$1\,kN/m stiffness). A 1\,mm horizontally shifted 100\,mg mass causes $1\,\mu$N differential load and thereby induces approximately 1\,nrad tilt. In conjunction with the above mentioned tilt to yaw coupling this can result on the order of 1\,nm test mass motion per mm of 100\,mg source mass motion, which is on the order of the observed signal and dominating the recoil effect. In practice, the air springs employ active tilt feedback at low frequencies, stiffening the table tilt motion. When deflated, the table leg stiffness may increase to 1\,MN/m, providing a factor 1000 suppression of source mass induced table tilt. Internal table flexing modes even in the low 100\,Hz region would not allow for increased tilt coupling at the modulation frequency. For the future a more in depth investigation of tilt to yaw coupling and cancellation of both effects by means of a counter-acting mass will be investigated.

Therefore, both effects are indeed of importance for massive source masses and high accuracy measurements aimed to determine $G$ very accurately. But to our current understanding they can be neglected in the presented experiment and similar future low frequency designs aiming for further reduced source masses.
        
\section{Magnetic interaction}\label{sec:magnetic}
To minimize influence from magnetic coupling the experiment was designed without ferromagnetic materials. All masses are made of diamagnetic gold, and the source mass is mounted on a paramagnetic titanium rod. As a consequence, Earth's magnetic field ($\approx50\,\mathrm{\mu T}$ at our location) as well as anthropogenic magnetic noise in an urban environment induce dipoles in the gold spheres leading to a repulsive force between them.
        
Using literature values\cite{lide2004crc} for the permeability of commercial grade gold and titanium we estimate dipole forces 
\begin{equation}
    F_{mag} = \frac{6\mu_0m_1m_2}{4 \pi d_c^4}
    \label{eq:magnetic_force}
\end{equation}
between two magnetic dipoles $m_i$ at center distance $d_c$ such as between our two diamagnetic spheres as well as the force between the titanium rod and one gold sphere.
The strength of the dipoles is determined by the strength of the externally applied field, the susceptibility and the volume and geometry of the material.

Iron impurities in the gold are suggested to dominate the magnetic susceptibility\cite{shih1931magnetic}. Yet gold-iron alloys will act diamagnetically up to 1000\,ppm iron, paramagnetically up to 10\,\% iron and only then start acting as a ferromagnetic material.
The gold used in the experiment is specified to have impurities below 1000\,ppm, but according to Ref.~\cite{henry1956xxi} commercial grade gold has iron impurities between 10 and 100\,ppm.
As a rough experimental verification for the permeability, a $1\,\mathrm{cm}$ cube NeFeB magnet with a remanent magnetization of around $1.4\,\mathrm{T}$ was brought within $7\,\mathrm{mm}$ distance of test and feedback mass. The literature value of $X_{Au}$ is confirmed within a factor of $\approx4$.

Projecting the dipoles induced by earth's magnetic field to exerted forces by means of Equation \ref{eq:magnetic_force}, they are about $4$ orders of magnitudes weaker than gravity in this setup.

\begin{figure}[htb]
\centering
\includegraphics[width=1.0\linewidth]{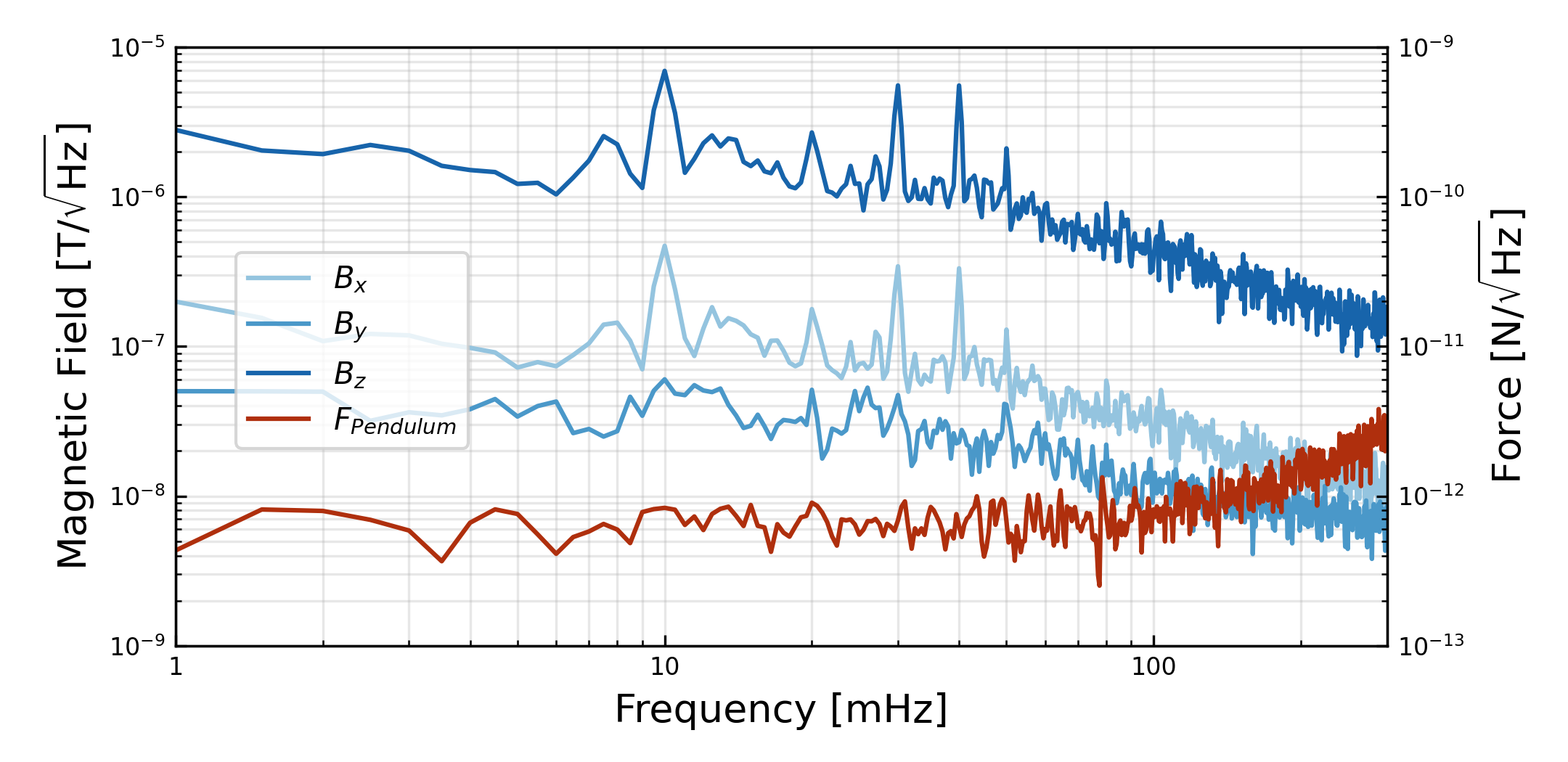}
\caption[magnetic coupling]{The amplitude spectral density of magnetic field variations at the site of the experiment were measured with a 3-axis magnetometer \textit{Stefan Mayer FLC3-70} (z = vertical, y = along source-test mass axis). The signal at $10\,\mathrm{mHz}$ is explained by the period of the traffic light at a nearby crossing, regulating car traffic as well as trams. The inferred force acting on the pendulum (red line), shows no sign of this signal.}
\label{fig:magnetic}
\end{figure}

To estimate magnetic signals induced by anthropogenic fields we measured the magnetic field close to the experiment with a 3-axis magnetic sensor (\textit{Stefan Mayer FLC3-70}), and two different \textit{Samsung} phones containing fluxgate magnetometers (see Figure \ref{fig:magnetic}).
The $10\,\mathrm{mHz}$ frequency comb is correlated with a close-by traffic light period.
We verified that it is a magnetic signal by measuring with different magnetometers using different sampling ADCs.
Moreover, the signal strength does not depend on the exact location in the lab.
Yet, it is not visible in the force signal of the experiment. We therefore conclude that the masses do not couple magnetically via dipoles induced from anthropogenic magnetism.

\section{Casimir force}\label{sec:Casimir}
The steady state Casimir force for a plane-sphere geometry can be calculated\cite{sushkov2011observation} using
\begin{equation}
    F_{Casimir} = \frac{r_t k_B T \zeta(3)}{8 d_s^2}.
\end{equation}
where $\zeta$ is the Riemann zeta function.
For our setup, the parameters are $T=300\,\mathrm{K}$, $r_t = 1\,\mathrm{mm}$ and $d_s\approx 250\,\mathrm{\mu m}$.
Thus, the Casimir force between the test mass and the Faraday shield is more than four orders of magnitude smaller than Newtonian gravity\cite{Emig2007}.
For extending the experiment towards smaller source masses, however, the Casimir force will dominate the gravitational attraction of a Planck mass sized source object for shield-sphere separations below $\approx$ 100\,$\mu$m.
As in our approach the gravitational potential is modulated, even in that source mass regime we will be able to distinguish Casimir forces exerted onto the test mass (which are static in the presence of an electrostatic shield) from gravity. Furthermore, for reduced shield separations the Casimir force might become relevant as a static non-linearity or instability of the torsion balance potential. For a 183\,mg oscillator with a natural frequency of 1\,mHz the Casimir force will render the oscillator unstable by surpassing the restoring force only for surface distances below $\approx100$\,nm.

For any set up without a Faraday shield Casimir forces set a lower bound for test mass to source mass separation which in turn sets an ultimate lower limit on detectable Newtonian force. For our scheme these limits are approximately 2\,$\mu$m for the symmetric setup with 90\,mg masses and 20\,$\mu$m for a source object with $m=m_{Planck}$.

\section{Systematic deviations}
In this section we list identified sources of systematic over- or underestimation of the measured gravitational force. Neither of these has been corrected for in the evaluation as most can only be guessed. We present Table II to assess the most relevant contributions considered. In general, we find that most effects scale with the source mass. Hence, they limit the accuracy of our $G$-estimation but will not pose a limit to measure smaller source masses in future.
        
All involved masses except for the tiny amounts of glue used have been measured at a 0.1\,mg accuracy (\textit{Kern ABS120-4}).
While in the evaluation we treated our source and test mass as point sources of gravity, they also act gravitationally onto the other components such as capillary, glue etc. as well. 
Gravity between the source mass support rod (titanium gravity) and the test mass is numerically simulated by integrating over $(10\,\mathrm{\mu m})^3$ elements. Since its mass distribution deviates from spherical symmetry it cannot be summarized as effective point-mass co-moving with the source mass. Instead, the modulation caused over a range of $2.4-6$\,mm center of mass separation is put into relation to the gravity modulation caused by source/test mass modulation over the same range.
Similarly, the influence of the spatial extent of the hole for mounting the titanium rod inside the source mass is estimated. The difference between a spherically symmetric source mass with a $300\,\mathrm{\mu m}$ hole and a spherically symmetric source mass without a hole but with the same mass is evaluated and put into relation with nominal source/test mass gravity.
The modulation of gravity between the counterbalance mass of the torsion pendulum and the source mass is quantified over the same $2.4-6$\,mm range.
The calibration of the QPD signal was done once on best effort basis, results are shown in Figure \ref{fig:PDcalibration}. The quality was checked by varying assumptions, but cannot be tested independently or continuously.
The source drive position is calibrated to a 1 pixel accuracy ($\approx20\,\mathrm{\mu m}$) at the beginning and end of each measurement run. The accuracy of the interpolation in between cannot be quantified as the camera could not be operated continuously.
Instead of only a 2D projection the 3D mass separation will be monitored continuously in future, improving the data for close approaches such as in Figure \ref{fig:Lekner-bonfire} by giving height offset and high frequency pendulum and roll mode excitation information. The 3D mass distribution of our gold 'spheres' is not known as we can only observe 2D projections. From microscopic imaging under different angles the used masses are known to be rather bumpy (source mass radii vary between 1.03 and 1.11\,mm) while in high accuracy experiments aiming for the determination of G even tiniest shape and density deviations are known to be relevant. 
\begin{center}
\begin{table}[ht]
\centering
            \begin{tabular}{l@{\hskip 3em}r}
                \toprule
                \textbf{effect } & \textbf{influence / $F_G$}\\
                \toprule
                hole gravity & +8.4E-3 \\[.25em]
                titanium gravity & +6.1E-3 \\[.25em]
                capillary gravity & +4.5E-3 \\[.25em]
                glue gravity & +3E-3 \\[.25em]
                counterbalance grav. & $-$1.1E-4 \\[.25em]
                electrostatic & $\pm$3E-2\\[.25em]
                magnetic & $\pm$1E-4\\[.25em]
                calibration & $\pm$6E-3\\[.25em]
                drive calibration & $\pm$1.6E-2\\[.25em]
                mass separation & $\pm$1E-3\\[.25em]
                $m_s$ accuracy & $\pm$1.1E-3\\[.25em]
                $m_t$ accuracy & $\pm$1.1E-3\\[.25em]
                $m_{glue}$ accuracy & +3E-3\\[.25em]
                height offset & $\pm$1.5E-2\\[.25em]
                $m_s$ roundness & $\pm$3.2E-2\\[.25em]
                $m_t$ roundness & $\pm$1.0E-2\\[.25em]
                $Q$ accuracy & $\pm$5E-3\\[.25em]
                bandpass $1f_m$ & +1.6E-2 \\[.25em]
                \textcolor{gray}{bandpass $2f_m$} & \textcolor{gray}{-7.6E-2} \\[.25em]
                downsampling & $\pm$1E-3\\[.25em]
                compensation pole & $\pm$1E-5\\[.25em]
                anti aliasing filter & $\pm$2E-6\\[.25em]
                \midrule
                upper limit & +15.9E-2\\[.25em]
                lower limit & $-$11.8E-2\\
                \bottomrule
            \end{tabular}
\label{tab:systematics}
\caption[Identified systemcatics]{identified systematic deviations in units of expected Newtonian gravity between source and test mass.}

\end{table}
\end{center}

\begin{figure}[htb]
    \centering\label{fig:shape}
    \includegraphics[width=\linewidth]{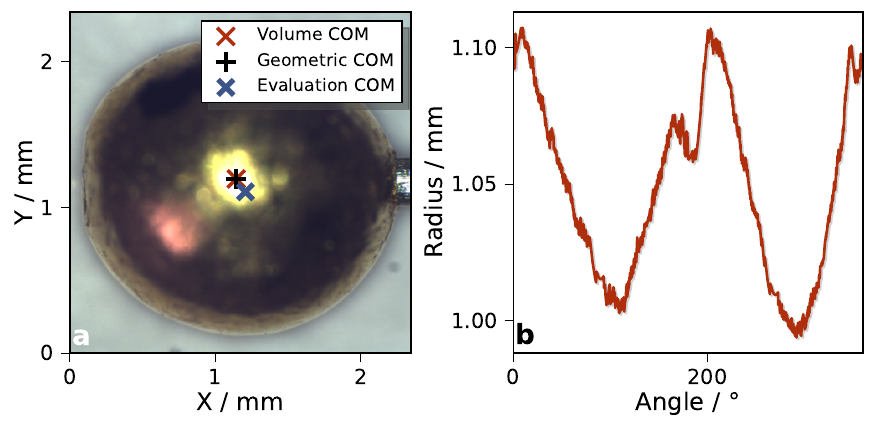}
    \caption[Shape of test mass sphere]{(a) 2D projection of the test mass: The center of mass determined by the geometric center of mass (mean over all image dimensions) as well as the center of mass calculated by our shape evaluation are indicated (Volume COM). The center of mass used in the determination of $G$ is indicated a blue cross. (b) The radius measured from the geometric COM over all angles is shown.}
    
\end{figure}

The masses used in the experiment deviate from perfect spheres. In order to quantify the influence of the shape deviations, we took photographs of the test sphere under a microscope and different angles as shown in Figure 13. We extract the angle resolved radius from the geometric center of mass of each projection. We assume, that each combination of angle and radius defines a spherical wedge and calculate the 2D deviation of the combined center of masses of the wedges from the geometric center of mass of the projection. We repeat the evaluation for every projection and take the mean deviation from the geometric center of mass as a correction for the center of mass distance used in the evaluation. Comparing fits to the uncorrected and corrected center of mass distance results in an estimated systematic uncertainty of $4.2\,\%$ for both, test and source mass shape. 
        
An additional damping rate caused by residual gas molecule impacts was estimated according to \cite{Schmoele2017}. Using our experimental parameters and taking into account the close proximity of the EM-shield to the test mass, we obtain a damping rate 5 orders of magnitude smaller than the natural damping rate of our test mass oscillator.

A worst case estimate for sorption effects on uncleaned gold surfaces \cite{Beer_2002, Glaeser_2009}, results in a relative mass change in the order of $10^{-7}$ and can therefore be neglected. We also note that we do not observe time-dependent systematic effects in agreement with the absence of significant sorption effects.
\end{document}